\documentclass{mn2e}
\pdfoutput=1 

\usepackage{epsf}
\usepackage[utf8]{inputenc}
\usepackage{amstext}
\usepackage{graphicx}
\usepackage{amssymb}
\usepackage{amsmath}
\usepackage{multirow}
\usepackage{color}

\newcommand{\be}{\begin{equation}}
\newcommand{\ee}{\end{equation}}
\def\beq{\begin{equation}}
\def\eeq{\end{equation}}
\def\ber{\begin{eqnarray}}
\def\eer{\end{eqnarray}}

\newcommand{\rma}{\rho_m}

\newcommand{\rL}{\rho_\Lambda}

\newcommand{\CC}{\Lambda}

\newcommand{\OL}{\Omega_{\Lambda}}

\newcommand{\rco}{\rho_{c 0}}

\newcommand{\rmr}{\rho_m}

\newcommand{\rLo}{\rho_{\CC 0}}

\begin{document}
\title[Possible signals of vacuum dynamics in the Universe]{Possible signals of vacuum dynamics in the Universe}
\author[Sol\`a Peracaula, de Cruz P\'erez and G\'omez-Valent]
{Joan Sol\`a Peracaula\thanks{sola@fqa.ub.edu}, Javier de Cruz P\'erez\thanks{decruz@fqa.ub.edu}, Adri\`a G\'omez-Valent\thanks{adriagova@fqa.ub.edu}
 \\
Departament  de F\'\i sica Qu\`antica i Astrof\'\i sica, and Institute of Cosmos Sciences, Universitat de Barcelona, Av. Diagonal 647\\
\vspace{0.1cm}
 E-08028 Barcelona, Catalonia, Spain
}

\date{\today}

\maketitle

\begin{abstract}
We study a generic class of time-evolving vacuum models which can provide a  better phenomenological account of the overall  cosmological observations as compared to the $\CC$CDM. Among these models, the running vacuum model (RVM) appears to be the most motivated and favored one, at a confidence level of $\sim 3\sigma$. We further support these results by computing the Akaike and Bayesian information criteria.  Our analysis also shows that we can extract fair signals of dynamical dark energy (DDE) by confronting the same set of data to the generic XCDM and CPL parametrizations. In all cases we confirm that the combined triad of modern observations on Baryonic Acoustic Oscillations, Large Scale Structure formation, and the Cosmic Microwave Background, provide the bulk of the signal sustaining a  possible vacuum dynamics. In the absence of any of these three crucial data sources, the DDE signal can not be perceived at a significant confidence level. Its possible existence could be a cure for some of the tensions existing in the $\CC$CDM when confronted to observations.
\end{abstract}


\section{Introduction}
Observations over the years seem to firmly support the current acceleration of the universe and therefore the possible existence of a generic cause responsible for it which we call dark energy (DE) see e.g. (Riess et al. 1998; Perlmutter et al. 1999; WMAP collab. 2013; Planck collab. XVI 2014; Planck collab. XIII 2016; Planck collab. XIV 2016) and references therein. Cosmologists have worked hard to decipher the dark energy code, but we still ignore the physical nature of the DE and hence of the ultimate cause of the observed acceleration of the universe. Such theoretical conundrum is the so-called Cosmological Constant Problem (CCP) (Weinberg 1989; Sahni \& Starobinsky 2000; Padmanabhan 2003; Peebles \& Ratra 2003; Copeland, Sami \& Tsujikawa 2006; Sol\`a 2013). In fact, the  cosmological constant (CC), $\CC$, or equivalently the vacuum energy density associated to it, $\rL=\CC/(8\pi G)$ ($G$ being Newton's gravitational coupling), is usually regarded as the simplest possible explanation for the DE.
Historically, the CC was introduced by Einstein in the gravitational field equations one hundred and one years ago (Einstein 1917). A positive, constant, tiny value (in particle physics units) of order $\rL\sim 2.7\times 10^{-47}$ GeV$^4\sim \left(2.3\times 10^{-3}{\rm eV}\right)^4$  can explain the needed speed up of our cosmos according to the observations.  The standard or ``concordance'' cosmological model embodies such an assumption as a fundamental built-in principle, together with the hypothesis of dark matter (DM), and for this reason is called the $\Lambda$CDM model. Formulated in terms of the current cosmological parameters, the $\Lambda$CDM assumes that $\rL=$const. throughout the history of the universe, with $\Omega_\CC\simeq 0.7$ and $\Omega_m\simeq 0.3$ at present.  Unfortunately, no convincing theoretical explanation is provided about the measured value of $\rL$. At the end of the day, no fundamental theory, not even quantum field theory (QFT), can explain this value; and, what is worse, the typical prediction is preposterously large as compared to the measured value. The difficulties inherent to this concept were recognized as of the time when Y.B. Zeldovich first observed (Zeldovich 1967) that the contribution from QFT to the vacuum energy density should be of order  $\sim m^4$ for any quantum field of mass $m$, and therefore many orders of magnitude bigger than the existing upper bound on $\rL$ in those days.

Since long  cosmologists have  felt motivated to look for alternative explanations for the DE beyond a rigid cosmological constant $\CC$. The scalar field paradigm was then profusely used also to make the cosmic vacuum dynamical: $\CC=\CC(\phi)$. In the old days the main aim was to adjust the large value of $\CC$ typically predicted in QFT to be zero. There were many early proposals, see e.g. (Endo \& Fukui 1977, 1982; Fujii 1982; Dolgov 1983; Abbott 1985; Zee 1985; Barr 1987; Ford 1987; Peccei, Sol\`a \& Wetterich 1987; Weiss 1987; Barr \& Hochberg 1988). In spite of the hopes raised by these works at solving the ``old CC Problem'', it was later shown in (Weinberg 1989) through the so-called ``no-go theorem'' that most if not all the dynamical adjustment mechanisms existing in the literature to date were plagued by  more or less obvious forms of subtly hidden fine tuning. For this reason the subsequent use of scalar fields in cosmology was mostly focused on trying to explain another aspect of the CCP: the cosmic coincidence problem (viz. the fact that $\rL$ happens to be so close to the matter density $\rma$ right now), see e.g. (Peebles \& Ratra 2003). The new wave of dynamical scalar fields in cosmology crystalized in the notions of quintessence, phantom fields and the like, which have had a tremendous influence in cosmology till our days: see e.g. (Peebles \& Ratra 1988; Ratra \& Peebles 1988; Wetterich 1988; Wetterich 1995; Caldwell, Dave \& Steinhardt 1998; Zlatev, Wang \& Steinhardt 1999; Amendola 2000; Caldwell, Kamionkowski \& Weinberg 2003), the reviews (Sahni \& Starobinsky 2000; Padmanabhan 2003; Peebles \& Ratra 2003; Copeland, Sami \& Tsujikawa 2006) and the many references therein. At the same time a blooming crest of models based on ascribing a direct phenomenological time-dependence to the CC term, $\CC=\CC(t)$, broke with impetus into the market. For an account of some of the old attempts, see (Overduin \& Cooperstock 1998) and references therein.

In this work, rather than attempting to solve the underlying theoretical  enigmas affecting the $\CC$CDM we wish to address more practical matters . We wish to follow the original phenomenological approach that made possible to unveil that $\rL$ is nonvanishing, irrespective of its ultimate physical nature. The method was largely empirical, namely $\rL$ was assumed to be a parameter and then fitted directly to the data. Of course a minimal set of assumptions had to be made, such as the validity of the Cosmological Principle and hence of the Friedmann-Lema\^itre-Robertson-Walker (FLRW) metric, with the ensuing set of Friedmann equations for the scale factor (Peebles 1993).  In our case, we wish of course to keep these minimal assumptions and make a phenomenological case study of the possibility that $\CC$ might be not just a parameter but a slowly varying cosmic variable mimicking the $\CC$CDM-like behavior. Furthermore, we motivate our study by considering the possibility that the inherent dynamics in $\rL$ is connected to fundamental aspects of QFT. In fact, within the class of dynamical vacuum models (DVMs), one of the main models under study is the ``running vacuum model'' (RVM), which can be connected to important aspects of  QFT in curved spacetime, see (Sol\`a 2013; Sol\`a \& G\'omez-Valent 2015; Sol\`a 2016; Sol\`a 2008) and references therein. One can think of this framework as one in which the $\CC$CDM is replaced by $\bar{\CC}$CDM (Sol\`a \& G\'omez-Valent 2015), with $\bar{\CC}={\CC}(H)$, or equivalently $\rL=\rL(H)$, playing the role of ``running''vacuum energy density. Interestingly, such a running with the expansion rate, $H$, can be related to the renormalization group. For previous investigations along these lines, see e.g. ( Espa\~na-Bonet et al, 2004, 2003; Babi\'{c}, Guberina, Horvat \& \v{S}tefan\v{c}i\'{c} 2005; Basilakos, Plionis \& Sol\`a 2009; Sol\`a 2011; Grande, So\`a, Basilakos \& Plionis 2011; Basilakos, Polarski \& Sol\`a 2012; Basilakos \& Sol\`a 2014; G\'omez-Valent, Sol\`a \& Basilakos 2015; G\'omez-Valent \& Sol\`a 2015; Sol\`a, G\'omez-Valent \& de Cruz P\'erez 2015; G\'omez-Valent, Karimkhani \& Sol\`a 2015; Basilakos 2015; Geng, Lee \& Zhang 2016; Geng, Lee \& Yin 2017), and the closely related recent works (Sol\`a, G\'omez-Valent \& de Cruz P\'erez 2017a,b,c,d and Sol\`a, de Cruz P\'erez  \& G\'omez-Valent  2018). It turns out that the peak confidence level  for DDE that we find in the context of the DVMs is near $\gtrsim 3.5\sigma$ at present. Interestingly, when we confront the same data with a simple XCDM (Turner \& White 1997) or CPL (Chevallier \& Polarski 2001; Linder 2003, 2004) parametrizations of the DDE (Amendola \& Tsujikawa 2015) we can still extract significant evidence of vacuum dynamics, showing that the signal is not restricted to particular models but it is rather generic. The first relatively recent signs of such dynamics were advanced in (G\'omez-Valent, Sol\`a \& Basilakos 2015; G\'omez-Valent \& Sol\`a 2015; Sol\`a, G\'omez-Valent \& de Cruz P\'erez 2015; G\'omez-Valent, Karimkhani \& Sol\`a 2015). Since then new support to the dynamical DE from the observational point of view has appeared in the literature using different methods and attaining a similar confidence level (Zhao G-B et al. 2017).

Finally, in view of the practical nature of the present work we keep an eye to the fact that the $\CC$CDM is afflicted of several persistent tensions when compared to the cosmological data. Such tensions involve relevant parameters of cosmology, such as the Hubble parameter, i.e. the current value of the Hubble function, $H(t_0)\equiv H_0$, and the current value of the rms of mass fluctuations at spheres of $8\,h^{-1}$ Mpc, i.e.  $\sigma_8(0)$. Such situation could be caused by as-yet unrecognized uncertainties or hint at physics beyond the $\CC$CDM (Freedman 2017). We cannot exclude e.g. that the peculiarities of important cosmological processes, for instance those related with structure formation, are conspicuously sensitive and even positively receptive, to a mild dynamical variation of the cosmic vacuum, which certainly influences the gravitational interaction of matter. Recall that a positive $\CC$ suppresses the growth of structure formation and this explains why the $\CC$CDM model is highly preferred to the CDM one, in which $\CC=0$.  Therefore, it is natural to reconsider these processes by considering the effect of a time modulation of the growth suppression through $\rL=\rL(t)$. We find that this helps to ameliorate the description of the  large scale structure (LSS) formation data, so it is worthwhile testing it.  For some studies addressing the existing tensions from various perspectives see e.g. (Valentino, Melchiorri, Linder \& Silk 2017; Valentino, Melchiorri \& Mena 2017;  Zhai et al., 2017; Sol\`a, G\'omez-Valent \& de Cruz P\'erez 2017c,d; G\'omez-Valent \& Sol\`a 2017, 2018).  See also (Chen Y. et al., 2016; Yu, Ratra \& Wang 2018).

The guidelines of our work are as follows. In Sect.\,\ref{sect:DVMs} we describe the dynamical vacuum models (DVMs). In Sect.\,\ref{sect:Fit} we report on the set of cosmological data used, on distant type Ia supernovae (SNIa), baryonic acoustic oscillations (BAOs), the Hubble parameter values at different redshifts, the LSS data, and the cosmic microwave background (CMB) from Planck. In Sect.\,\ref{sect:perturbationsDVM} we discuss aspects of structure formation with dynamical vacuum. The numerical analysis of the DVMs and a comparison with the standard XCDM and CPL parametrizations is the object of Sect.\,\ref{sect:numerical results}. An ample discussion of the results along with a reanalysis under different conditions is developed in Sect.\,\ref{sect:discussion}. Finally, in Sect.\,\ref{sect:conclusions} we present our conclusions.

\section{Dynamical vacuum models}\label{sect:DVMs}
{The gravitational field equations  are $G_{\mu\nu}=8\pi G\ \tilde{T}_{\mu\nu}$,
where $G_{\mu\nu}=R_{\mu \nu }-\frac{1}{2}g_{\mu \nu }R$ is the Einstein tensor and
$\tilde{T}_{\mu\nu}\equiv T_{\mu\nu}+g_{\mu\nu}\,\rL $ is the full energy-momentum tensor involving the effect of both matter and the vacuum energy density, with $\rL=\CC/(8\pi G)$.
The structure of $\tilde{T}_{\mu\nu}$ shows that the vacuum is dealt with as a perfect fluid carrying an equation of state (EoS)  $p_{\CC}=-\rho_{\CC}$. When the matter can also be treated as an ideal fluid and is distributed homogeneously and isotropically, as postulated by the Cosmological Principle, we can write
$\tilde{T}_{\mu\nu}= (\rho_{\Lambda}-p_{m})\,g_{\mu\nu}+(\rho_{m}+p_{m})U_{\mu}U_{\nu}$,
where $U_{\mu}$ is the bulk $4$-velocity of the cosmic fluid, $\rho_m$ is the proper energy density of matter and $p_m$ its isotropic pressure.}
We assume the standard cosmological framework grounded on the FLRW metric with flat three-dimensional slices: $ds^2=dt^2-a^2(t)\,d{\bf x}^2$, where $t$ is the cosmic time and $a(t)$ the scale factor. However, we admit that matter can be in interaction with vacuum, which is tantamount to saying that $\rL=\rL(\zeta)$ is a function of some cosmic variable evolving with time, $\zeta=\zeta(t)$. While this, of course, implies that $\dot{\rho}_{\CC}\equiv d\rL/dt\neq 0$ we assume that $\dot{G}=0$ in our study -- see (Sol\`a, G\'omez-Valent \& de Cruz P\'erez 2015, 2017a) for studies including the option $\dot{G}\neq 0$. Such  vacuum dynamics is compatible with the Bianchi identity (see below) provided there is some energy exchange between vacuum and matter. It means that matter cannot be strictly conserved in these circumstances. The standard Friedmann and acceleration equations for the present universe remain formally identical to the standard $\CC$CDM case:
\begin{eqnarray}
&&3H^2=8\pi\,G\sum_N\rho_N=8\pi\,G\,(\rho_m+\rho_r+\rho_\Lambda(\zeta))\label{eq:FriedmannEq}\\
&&3H^2+2\dot{H}=-8\pi\,G\sum_Np_N=-8\pi\,G\,(p_r-\rho_\Lambda(\zeta))\label{eq:PressureEq}\,.
\end{eqnarray}
Here $H=\dot{a}/a$ is the usual Hubble function,  $\rho_m=\rho_b+\rho_{dm}$ involves the pressureless contributions from baryons and cold DM, and $\rho_r$ is  the radiation density (with the usual EoS $p_r=\rho_r/3$).  We emphasize once more that in the above equations we stick to the EoS $p_{\CC}=-\rho_{\CC}$, although the vacuum is dynamical, $\rL(t)=\rL(\zeta(t))$, and its evolution is tied to the cosmic expansion. {The sums above run over all the components $N=dm,b,r,\CC$}.  In all of the dynamical vacuum models (DVMs) being considered here, the cosmic variable $\zeta$ is either the scale factor or can be expressed analytically in terms of it, $\zeta=\zeta(a)$, or equivalently in terms of the cosmological redshift, $z=a^{-1}-1$, in which we adopt the normalization $a=1$ at present.
{From the basic pair of equations \eqref{eq:FriedmannEq}-\eqref{eq:PressureEq}, a first integral of the system follows:

\begin{equation}
\begin{split}
\sum_{N} \dot{\rho}_N &+3\,H(\rho_N+p_N)= \label{BianchiGeneral}\\
& \dot{\rho}_\CC +\dot{\rho}_{dm} + 3H\rho_{dm}+\dot{\rho}_b + 3H\rho_b+\dot{\rho}_r + 4H\rho_r =0\,.
\end{split}
\end{equation}
Such a first integral ensues also from the divergenceless property of the full energy-momentum tensor $\tilde{T}_{\mu\nu}$ in the FLRW metric, i.e. $\nabla^{\mu}\tilde{T}_{\mu\nu}=0$. The last property is a consequence of the Bianchi identity satisfied by the Einstein tensor, $\nabla^{\mu} G_{\mu\nu}=0$, and the assumed constancy of the Newtonian coupling $G$. It reflects the local conservation law of the compound system formed by matter and vacuum, and the consequent nonconservation of each of these components when taken separately.}

The concordance model assumes that matter and radiation are self-conserved after equality. It also assumes that baryons and CDM are separately conserved. Hence their respective energy densities satisfy $\dot{\rho}_b + 3H\rho_b=0$, $\dot{\rho}_r + 4H\rho_r=0$ and $\dot{\rho}_{dm} + 3H\rho_{dm}=0$.  In the presence of vacuum dynamics it is obvious that at least one of these equations cannot hold. Following our definite purpose to remain as close as possible to the $\CC$CDM, we shall assume that the first two of the mentioned conservation equations still hold good but that the last does not, meaning that the vacuum exchanges energy only with DM. The dilution laws for baryons and radiation as a function of the scale factor therefore take on the conventional $\CC$CDM forms:
\begin{equation}\label{eq:BaryonsRadiation}
\rho_b(a) = \rho_{b0}\,a^{-3}, \ \ \ \ \ \ \ \rho_r(a)=\rho_{r0}\,a^{-4}\,,
\end{equation}
where $\rho_{b0}$ and $\rho_{r0}$ are the corresponding current values. In contrast, the density of DM is tied to the dynamics of the vacuum. {Taking into account the conserved components and  introducing the vacuum-dark matter interaction source, $Q$, we can write the interactive part of (\ref{BianchiGeneral}) into two coupled equations:}
\begin{equation}\label{eq:Qequations}
\dot{\rho}_{dm}+3H\rho_{dm}=Q\,,\ \ \ \ \ \ \ \dot\rho_{\CC}=-{Q}\,.
\end{equation}
The solution of these equations will depend on the particular form assumed for $Q$, which determines the leakage rate of vacuum energy into dark matter or vice versa. Such a leakage must certainly be much smaller than the standard dilution rate of nonrelativistic matter associated to the cosmic expansion (i.e. much smaller than $\sim a^{-3}$), as otherwise these anomalous effects would be too sharp at the present time. Therefore, we must have   $0<|Q|\ll\dot{\rho}_m$. The different DVMs will be characterized by different functions $Q_i$  ($i=1,2,..$).

Two possible phenomenological ansatzs  considered in the literature  are (Salvatelli et al. 2014; Murgia, Gariazzo \& Fornengo 2016; Li, Zhang \& Zhang 2016; Zhao M-M, et al, 2017; Guo, Zhang \& Zhang 2018)
\begin{eqnarray}
{\ \rm Model\ \ }Q_{dm}: \phantom{XX}Q_{dm}&=&3\nu_{dm}H\rho_{dm}\label{eq:PhenModelQdm}\\
{\rm Model\ \ }Q_{\CC}:\phantom{XXx}Q_{\CC}&=&3\nu_{\CC}H\rho_{\CC}\,.\label{eq:PhenModelQL}
\end{eqnarray}
The dimensionless parameters $\nu_{i}=(\nu_{dm},\nu_\CC)$ for each model  ($Q_{dm}$, $Q_\CC$) determine the strength of the dark-sector interaction in the sources $Q_i$ and enable the evolution of the vacuum energy density.  For $\nu_{i}>0$ the vacuum decays into DM (which is thermodynamically favorable (Salim \& Waga 1996; Lima 1996)) whereas for $\nu_{i}<0$ is the other way around. This is also a relevant argument to judge the viability of these models, as only the first situation is compatible with the second law of thermodynamics.
There are many more choices for $Q$, see e.g. (Bolotin, Kostenko, Lemets \& Yerokhin 2015; Costa, Xu, Wang \& Abdalla 2017), but it will suffice to focus on these models and the RVM one defined in the next section to effectively assess the possible impact of the DVMs in the light of the modern observational data.

\subsection{The running vacuum model (RVM)}\label{sect:RVM}

The last DVM under study is the so-called running vacuum model (RVM), which can be motivated in the context of QFT in curved spacetime (cf. Sol\`a 2013; Sol\`a \& G\'omez-Valent 2015, and references therein).
The model has some virtues and can be extended to afford an effective description of the cosmic evolution starting from inflation up to our days (Sol\`a 2013; Sol\`a \& G\'omez-Valent 2015; Perico, Lima, Basilakos \& Sol\`a 2013; Lima, Basilakos \& Sol\`a 2013, 2015, 2016; Sol\`a 2015). For instance, in (Sol\`a 2015) it is suggested that the RVM could positively impinge on solving some of the fundamental cosmological problems, including the entropy problem. Intriguingly, the inherent tiny leakage of vacuum energy into matter within the RVM could also furnish an explanation for the possible slow time variation of the fundamental constants, an issue that has been examined in detail in (Fritzsch \& Sol\`a 2012, 2015; Fritzsch, Sol\`a \& Nunes 2017). See also the old work (Terazawa 1981). We shall not discuss here the implications for the early universe, but only for the part of the cosmic history that is accessible to our measurements and can therefore be tested phenomenologically with the current data.

As advertised, for the specific RVM case the cosmic variable $\zeta$ in the field equations (\ref{eq:FriedmannEq})-(\ref{eq:PressureEq}) can be identified with the Hubble rate $H$. The form of the RVM for the post-inflationary epoch and hence relevant for the current universe reads as follows:
\begin{equation}\label{eq:RVMvacuumdadensity}
\rho_\CC(H) = \frac{3}{8\pi{G}}\left(c_{0} + \nu{H^2}\right)\,.
\end{equation}
{Such structure  can be linked to a renormalization group (RG) equation for the vacuum energy density, in which the running scale $\mu$ of the RG is associated with the characteristic energy scale of the FLRW metric, i.e. $\mu=H$.
The additive constant $c_0=H_0^2\left(\Omega_\CC-\nu\right)$ appears because one integrates the RG equation satisfied by $\rL(H)$. It is fixed by the boundary condition $\rL(H_0)=\rLo$, where $\rLo$ and $H_0$ are the current values of these quantities; similarly $\OL=\rLo/\rco$  and  $\rco=3H_0^2/(8\pi G)$ are the values of the vacuum density parameter and the critical density today. The dimensionless coefficient $\nu$ encodes the dynamics of the vacuum at low energy and can be related with the $\beta$-function of the running of $\rL$.} Thus, we naturally expect $|\nu|\ll1$. An estimate of $\nu$ at one loop in QFT indicates that is of order $10^{-3}$ at most (Sol\`a 2008), but here we will treat it as a free parameter. This means  we shall deal with the RVM on pure phenomenological grounds, hence fitting actually $\nu$ to the observational data (cf. Sect.\,\ref{sect:Fit}).

In the RVM case, the source function $Q$ in \eqref{eq:Qequations} is not just put by hand (as in the case of the DVMs introduced before). It is a calculable expression from \eqref{eq:RVMvacuumdadensity}, using Friedmann's equation (\ref{eq:FriedmannEq}) and the local conservation laws (\ref{eq:BaryonsRadiation})-(\ref{eq:Qequations}). We find:
\begin{equation}\label{eq:QRVM}
{\rm RVM:}\qquad Q=-\dot{\rho}_{\Lambda}=\nu\,H(3\rho_{m}+4\rho_r)\,,
\end{equation}
where we recall that $\rho_{m}=\rho_{dm}+\rho_{b}$, and that $\rho_b$ and $\rho_r$ are known functions of the scale factor -- see Eq.\,(\ref{eq:BaryonsRadiation}). The remaining densities, $\rho_{dm}$ and $\rL$, must be determined upon further solving the model explicitly, see subsection\,\ref{sect:solvingDVM}. If baryons and radiation would also possess a small interaction with vacuum and/or $G$ would evolve with time, the cosmological solutions would be different (G\'omez-Valent, Sol\`a \& Basilakos 2015; G\'omez-Valent \& Sol\`a 2015; Basilakos 2015; Sol\`a, G\'omez-Valent \& de Cruz P\'erez 2015, 2017a). We can see from (\ref{eq:QRVM})  that the parameter $\nu$ plays a similar role as  $(\nu_{dm},\nu_\CC)$ for the more phenomenological models (\ref{eq:PhenModelQdm}) and (\ref{eq:PhenModelQL}). The three of them will be collectively denoted $\nu_i$.

\subsection{Solving explicitly the dynamical vacuum models}\label{sect:solvingDVM}

The matter and vacuum energy densities of the DVMs can be computed straightforwardly upon solving the coupled system of differential equations (\ref{eq:Qequations}), given the previous explicit forms for the interacting source in each case and keeping in mind that, in the current framework, the baryon ($\rho_b$) and radiation ($\rho_r$) parts are separately conserved. After some calculations the equations for the DM energy densities $\rho_{dm}$ for each model (RVM, $Q_{dm}$, $Q_{\CC}$) can be solved in terms of the scale factor. Below we quote the final results for each case:

\begin{eqnarray}\label{eq:rhoRVM}
{\rm \textbf{RVM}}:\quad \rho_{dm}(a) &=& \rho_{dm0}\,a^{-3(1-\nu)}+ \rho_{b0}\left(a^{-3(1-\nu)} - a^{-3}\right) \nonumber  \\
&+&
\frac{4\nu}{1 + 3\nu}\,{\rho_{r0}}\,\left(a^{-3(1-\nu)} - a^{-4}\right)
\end{eqnarray}
\begin{eqnarray}\label{eq:rhoQdm}
{\rm \ \mathbf{Q_{dm}}}:\quad
\rho_{dm}(a) = \rho_{dm0}\,a^{-3(1-\nu_{dm})}
\end{eqnarray}
\begin{eqnarray}\label{eq:rhoQL}
{\rm \ \mathbf{Q_{\CC}}}:\quad
\rho_{dm}(a) =\rho_{dm0}\,a^{-3} + \frac{\nu_\CC}{1-\nu_\CC}\rLo\left(a^{-3\nu_\Lambda}-a^{-3}\right)
\end{eqnarray}
%

\begin{small}
\begin{table*}
\begin{center}
\begin{scriptsize}
\resizebox{1\textwidth}{!}{
\begin{tabular}{| c | c |c | c | c | c | c | c | c | c | c|}
\multicolumn{1}{c}{Model} &  \multicolumn{1}{c}{$h$} &    \multicolumn{1}{c}{$\Omega_m$}&  \multicolumn{1}{c}{{\small$\nu_i$}}  & \multicolumn{1}{c}{$w_0$} & \multicolumn{1}{c}{$w_1$} &\multicolumn{1}{c}{$\sigma_8(0)$} & \multicolumn{1}{c}{$\Delta{\rm AIC}$} & \multicolumn{1}{c}{$\Delta{\rm BIC}$}\vspace{0.5mm}
\\\hline
{\small $\CC$CDM} & $0.692\pm 0.004$ &  $0.296\pm 0.004$ & - & -1 & - & $0.801\pm0.009$ & - & -\\
\hline
XCDM  &  $0.672\pm 0.007$&  $0.311\pm 0.007$& - & $-0.923\pm0.023$ & - & $0.767\pm0.014$ & 8.55 & 6.31 \\
\hline
CPL  &  $0.673\pm 0.009$&  $0.310\pm 0.009$& - & $-0.944\pm0.089$ & $0.063\pm0.259$ & $0.767\pm0.015$ & 6.30 & 1.87 \\
\hline
RVM  & $0.677\pm 0.005$&  $0.303\pm 0.005$ & $0.00158\pm 0.00042$ & -1 & - & $0.736\pm0.019$ & 12.91 & 10.67 \\
\hline
$Q_{dm}$ &  $0.678\pm 0.005$&  $0.302\pm 0.005 $ & $0.00216\pm 0.00060 $ & -1 & - &  $0.740\pm0.018$  & 12.13 & 9.89 \\
\hline
$Q_\CC$  &  $0.691\pm 0.004$&  $0.298\pm 0.005$ & $0.00601\pm 0.00253$ & -1 & - &  $0.790\pm0.010$ & 3.41 & 1.17 \\
\hline \end{tabular}
 }
\end{scriptsize}
\end{center}
\caption{Best-fit values for the $\CC$CDM, XCDM, CPL and the three dynamical vacuum models (DVMs).   {The specific fitting parameters for each DVM are $\nu_{i}=\nu $ (RVM), $\nu_{dm}$($Q_{dm}$) and $\nu_{\CC}$($Q_{\CC}$), whilst for the XCDM and CPL are the EoS parameters $w_0$ and the pair ($w_0$,$w_1$), respectively. For the DVMs and the $\CC$CDM, we have $w_0=-1$ and $w_1=0$. The remaining parameters are as in the $\CC$CDM and are not shown.  For convenience we reckon the values of $\sigma_8(0)$ for each model, which are not part of the fit but are computed from the fitted ones following the procedure indicated in Sect. 4.3 . The (positive) increments $\Delta$AIC and $\Delta$BIC (see the main text, Sect. 5.2) clearly favor the DDE options. The RVM and $Q_{dm}$ are particularly favored ($\sim 3.8\sigma$ c.l. and $3.6\sigma$, respectively). Our fit is performed over  a rich and fully updated SNIa+BAO+$H(z)$+LSS+CMB data set (cf. Sect. \ref{sect:Fit})}.}
\label{tableFit1}
\end{table*}
\end{small}

In solving the differential equations (\ref{eq:Qequations}) we have traded the cosmic time variable for the scale factor using the chain rule $d/dt=aH d/da$. The corresponding vacuum energy densities can also be solved in the same variable, and yield:

\begin{eqnarray}\label{eq:rhoVRVM}
{\rm \textbf{RVM}}:\quad \rho_\CC(a) &=& \rLo + \frac{\nu\,\rho_{m0}}{1-\nu}\left(a^{-3(1-\nu)}-1\right)\\&+&\frac{\nu{\rho_{r0}}}{1-\nu}\left(\frac{1-\nu}{1+3\nu}a^{-4} + \frac{4\nu}{1+3\nu}a^{-3(1-\nu)} -1\right)\nonumber
\end{eqnarray}
\begin{eqnarray}\label{eq:rhoVQdm}
{\rm \ \mathbf{Q_{dm}}}:\quad
\rho_\CC(a) = \rLo + \frac{\nu_{dm}\,\rho_{dm0}}{1-\nu_{dm}}\,\left(a^{-3(1-\nu_{dm})}-1\right)
\end{eqnarray}
\begin{eqnarray}\label{eq:rhoVQL}
{\rm \ \mathbf{Q_{\CC}}}:\quad
\rho_\CC(a) =\rLo\,{a^{-3\nu_\CC}}
\end{eqnarray}

One can easily check that for $a=1$ (i.e. at the present epoch) all of the energy densities (\ref{eq:rhoRVM})-(\ref{eq:rhoVQL}) recover their respective current values $\rho_{N0}$ ($N=dm,\CC$). In addition,
for $\nu_{i}\to 0$ we retrieve for the three DM densities the usual $\CC$CDM expression $\rho_{dm}(a)=\rho_{dm 0}a^{-3}$, and the corresponding vacuum energy densities $\rL(a)$ boil down to the constant value $\rLo$ in that limit. The normalized Hubble rate $E\equiv H/H_0$ for each model can be easily obtained by plugging the above formulas, along with the radiation and baryon energy densities \eqref{eq:BaryonsRadiation}, into  Friedmann's equation (\ref{eq:FriedmannEq}). We find:
\begin{eqnarray}\label{eq:HRVM}
{\rm \textbf{RVM}}:\quad E^2(a) &=& 1 + \frac{\Omega_m}{1-\nu}\left(a^{-3(1-\nu)}-1\right) \label{HRVM}\\ \nonumber\\
& +& \frac{\Omega_r}{1-\nu}\left(\frac{1-\nu}{1+3\nu}a^{-4} + \frac{4\nu}{1+3\nu}a^{-3(1-\nu)} -1\right)\nonumber
\end{eqnarray}
\begin{eqnarray}\label{HQdm}
{\rm \ \mathbf{Q_{dm}}}:\quad
E^2(a) &=& 1 + \Omega_b\left(a^{-3}-1\right) \\
&+& \frac{\Omega_{dm}}{1-\nu_{dm}}\left(a^{-3(1-\nu_{dm})}-1\right)
+ \Omega_r\left(a^{-4}-1\right)\nonumber
\end{eqnarray}

\begin{eqnarray}\label{HQL}
{\rm \ \mathbf{Q_{\CC}}}:\quad
E^2(a) &=&\frac{a^{-3\nu_\CC}-\nu_\CC{a^{-3}}}{1-\nu_\CC}+\frac{\Omega_m}{1-\nu_\CC}\left(a^{-3}-a^{-3\nu_\CC}\right) \nonumber\\ &+&  \Omega_r\left(a^{-4} + \frac{\nu_\CC}{1-\nu_\CC}a^{-3} - \frac{a^{-3\nu_\CC}}{1-\nu_\CC}\right)
\end{eqnarray}

In the above expressions, we have used the cosmological parameters $\Omega_N=\rho_{N0}/\rco$ for each fluid component ($N=dm,b,r,\CC$), and defined $\Omega_m=\Omega_{dm}+\Omega_b$. Altogether, they satisfy the sum rule $\sum_N\Omega_N=1$. The normalization condition $E(1)=1$  in these formulas is apparent, meaning that the Hubble function for each model reduces to $H_0$ at present, as they should; and, of course, for $\nu_i\to 0$ we recover the $\CC$CDM form for $H$, as should be expected.

\begin{figure*}
\centering
\includegraphics[angle=0,width=1.03\linewidth]{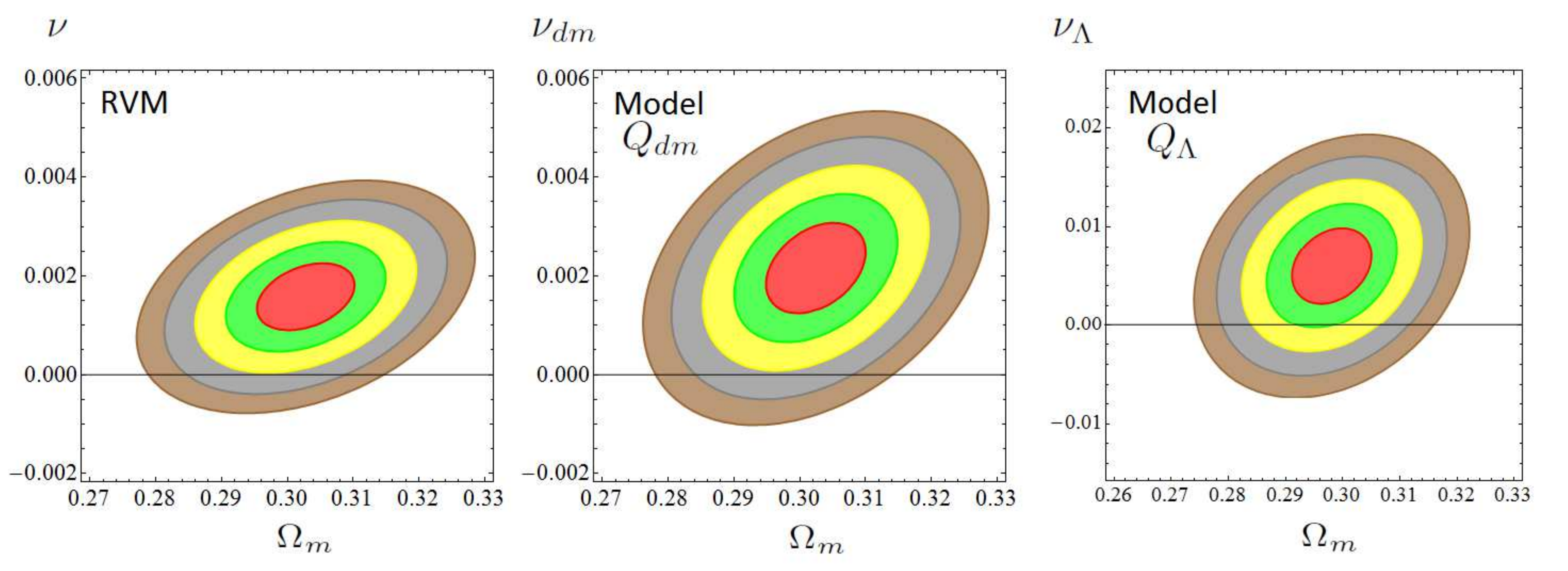}
\caption{\label{fig:G2Evolution}%
Likelihood contours for the DVMs in the $(\Omega_m,\nu_i)$ plane for the values $-2\ln\mathcal{L}/\mathcal{L}_{max}=2.30$, $6.18, 11.81$, $19.33$, $27.65$ (corresponding to 1$\sigma$, 2$\sigma$, 3$\sigma$, 4$\sigma$ and 5$\sigma$ c.l.) after marginalizing over the rest of the fitting parameters. We estimate that for the RVM, $94.80\%$ (resp. $89.16\%$) of the 4$\sigma$ (resp. 5$\sigma$) area is in the $\nu>0$ region. For the Q$_{dm}$ we find that $95.24\%$ (resp. $89.62\%$) of the 4$\sigma$ (resp. 5$\sigma$) area is in the $\nu_{dm}>0$ region. Finally, for the Q$_{\CC}$ we estimate that $99.45\%$ (resp. $90.22\%$) of the $2\sigma$ (resp. $3\sigma$) area is in the $\nu_{\CC}>0$ region. Subsequent marginalization over $\Omega_m$ increases slightly the c.l. and renders the fitting values collected in Table 1. The $\CC$CDM ($\nu_i=0$) appears disfavored at $\sim 4\sigma$ c.l. in the RVM and $Q_{dm}$, and at $\sim 2.5\sigma$ c.l. for $Q_\CC$.}
\end{figure*}
%

%
From the structure of equations (\ref{eq:rhoVRVM}) and (\ref{HRVM}) we can readily see that the vacuum energy density for the RVM can be fully written as a function of a cosmic variable $\zeta$, which can be chosen to be not only the scale factor but the full Hubble function $\zeta=H$. The result is, of course, Eq.\,(\ref{eq:RVMvacuumdadensity}).  In contrast, for the $Q_{dm}$ and $Q_{\CC}$ models this is not possible, as it is clear on comparing equations (\ref{eq:rhoVQdm})-(\ref{eq:rhoVQL}) and the corresponding ones (\ref{HQdm})-(\ref{HQL}). For these models $\rL$ can only be written as a function of the scale factor. Thus, the RVM happens to have the greatest level of symmetry since its origin is a RG equation in $H$ whose solution is precisely (\ref{eq:RVMvacuumdadensity}).

\subsection{XCDM and CPL parametrizations}\label{sect:XCDMandCPL}

Together with the DVMs , we fit also the same data through the simple XCDM parametrization of the dynamical DE, first introduced in (Turner \& White 1997). Since both matter and DE are self-conserved (i.e., they are not interacting), the DE density as a function of the scale factor is simply given by $\rho_X(a)=\rho_{X0}\,a^{-3(1+w_0)}$, with $\rho_{X0}=\rho_{\CC 0}$, where $w_0$ is the (constant) EoS parameter of  the generic DE entity X in this parametrization. The normalized Hubble function is:
\begin{equation}\label{eq:HXCDM}
E^2(a)=\Omega_m\,a^{-3}+\Omega_r\,a^{-4}+\OL\,a^{-3(1+w_0)}\,.
\end{equation}
For $w_0=-1$ it boils down to that of the $\CC$CDM with rigid CC term. The use of the XCDM parametrization throughout our analysis will be useful to roughly mimic a (noninteractive) DE scalar field with constant EoS. For $w_0\gtrsim-1$ the XCDM mimics quintessence, whereas for $w_0\lesssim-1$ it mimics phantom DE.

A slightly more sophisticated approximation to the behavior of a noninteractive scalar field playing the role of dynamical DE is afforded by the CPL parametrization (Chevallier, \& Polarski 2001; Linder 2003, 2004), in which one assumes that the generic DE entity $X$ has a slowly varying EoS of the form
\begin{equation}\label{eq:CPL}
w_D=w_0+w_1\,(1-a)=w_0+w_1\,\frac{z}{1+z}\,.
\end{equation}
The CPL parametrization, in contrast to the XCDM one, makes allowance for a time evolution of the dark energy EoS owing to the presence of the additional parameter $w_1$, which satisfies $0<|w_1|\ll|w_0|$, with $w_0\gtrsim -1$ or $w_0\lesssim -1$. The expression (\ref{eq:CPL}) is seen to have a well-defined limit both in the early universe ($a\to 0$, equivalently $z\to\infty$) and in the current one ($a=1$, or $z=0$).
The corresponding normalized Hubble function for the CPL can be easily found:
\begin{equation}
 E^2(a) = \Omega_m\,a^{-3}+ \Omega_r a^{-4}+\Omega_\CC
 a^{-3(1+w_0+w_1)}\,e^{-3\,w_1\,(1-a)}\,.
\label{eq:HCPL}
\end{equation}
The XCDM and the CPL parametrizations can be conceived as a kind of baseline frameworks to be referred to in the study of dynamical DE.  We expect that part of the DDE effects departing from the $\CC$CDM should be captured by these parametrizations, either in the form of effective quintessence behavior ($w\gtrsim -1$) or effective phantom behavior ($w\lesssim-1$).
The XCDM, though,  is the most appropriate for a fairer comparison with the DVMs, all of which also have one single vacuum parameter $\nu_i$.

\section{Data sets and results}\label{sect:Fit}

In this work, we fit the $\CC$CDM, XCDM, CPL and the three DVMs to the cosmological data from type Ia supernovae (Betoule et al. 2014), BAOs (Beutler et al. 2011; Kazin et al. 2014; Ross et al. 2015; Gil-Mar\'in et al. 2017; Delubac et al. 2015; Aubourg et al. 2015), the values of the Hubble parameter extracted from cosmic chronometers at various redshifts, $H(z_i)$ (Zhang et al. 2014, Jim\'enez et al. 2003; Simon, Verde \& Jim\'enez 2005; Moresco et al. 2012, 2016; Stern et al. 2010; Moresco 2015), the CMB data from Planck 2015 (Planck collab. XIII 2016) and the most updated set of LSS formation data embodied in the quantity $f(z_i)\sigma_8(z_i)$,
{see the corresponding values and references in Table 2}. It turns out that the LSS data is very important for the DDE signal, and up to some updating performed here it has been previously described in more detail in (Sol\`a, G\'omez-Valent \& de Cruz P\'erez 2017a). We denote this string of cosmological data by SNIa+BAO+$H(z)$+LSS+CMB.

A guide to the presentation of our results is the following. The various fitting analyses and contour plots under different conditions (to be discussed in detail in the next sections) are displayed in four fitting tables, Tables 1 and 3-5, and in seven figures, Figs.\,1-7. {The main numerical results of our analysis are those recorded in Table 1. Let us mention in particular  Fig. 6, whose aim is to  identify what are the main data responsible for the DDE effect under study.  Bearing in mind the aforementioned significance of the LSS data, Fig. 7 is aimed to compare in a graphical way the impact of the $f(z)\sigma_8(z)$ and weak lensing data on our results.
The remaining tables and figures contain complementary information, which can be helpful for a more detailed picture of our rather comprehensive study.
Worth noticing are the results displayed in Table 5, which shows what would be the outcome of our analysis if we would restrict ourselves to the fitting data employed by the Planck 2015 collaboration (Planck collab. XIII 2016), where e.g. no LSS data were used and no DDE signal was reported.  Additional  details and considerations are furnished of course in the rest of our exposition}.

\begin{figure*}
\centering
\includegraphics[width=15.0cm]{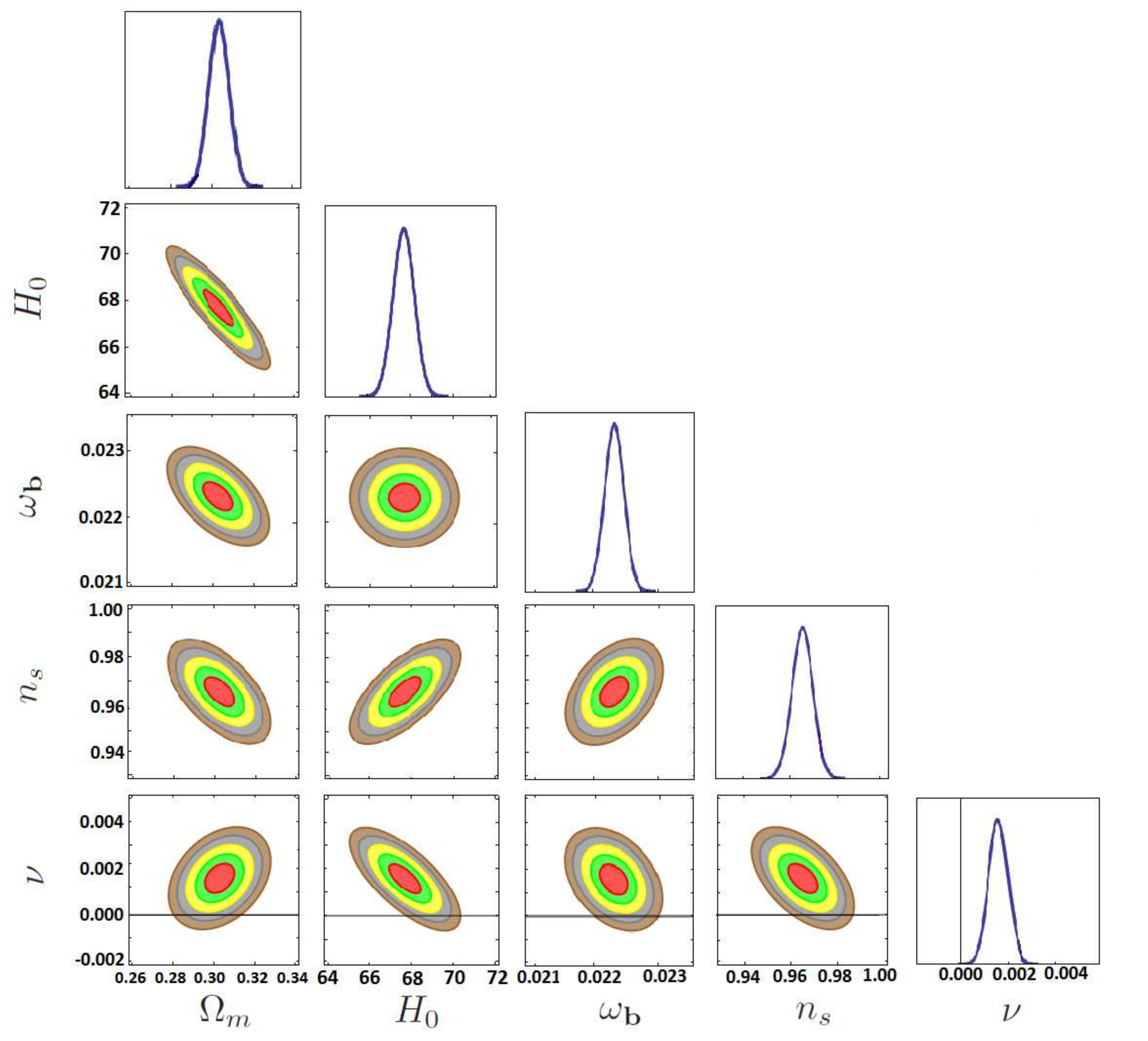}
\caption{\label{fig:MatrixPlot}%
As in Fig.\,1, but projecting the fitting results of the RVM onto the different planes of the involved parameters ($H_0$ is expressed in Km/s/Mpc). The horizontal line $\nu=0$ in the plots of the last row corresponds to the $\CC$CDM.   It is apparent that it is significantly excluded at $\sim 4\sigma$ c.l. in all cases. The peak in the rightmost plot corresponds to the central value $\nu=0.00158$ indicated in Table 1.
}
\end{figure*}
%

%
%
\begin{table}
\begin{center}
\resizebox{9cm}{!} {
\begin{tabular}{| c | c |c | c |}
\multicolumn{1}{c}{Survey} &  \multicolumn{1}{c}{$z$} &  \multicolumn{1}{c}{$f(z)\sigma_8(z)$} & \multicolumn{1}{c}{{\small References}}
\\\hline
6dFGS & $0.067$ & $0.423\pm 0.055$ & Beutler et al. 2012
\\\hline
SDSS-DR7 & $0.10$ & $0.37\pm 0.13$ & Feix, Nusser \& Branchini 2015
\\\hline
GAMA & $0.18$ & $0.29\pm 0.10$ & Simpson et al. 2016
\\ & $0.38$ & $0.44\pm0.06$ & Blake et al. 2013
\\\hline
DR12 BOSS & $0.32$ & $0.427\pm 0.056$  & Gil-Mar\'in et al. 2017\\
 & $0.57$ & $0.426\pm 0.029$ & \\\hline
WiggleZ & $0.22$ & $0.42\pm 0.07$ & Blake et al. 2011 \tabularnewline
 & $0.41$ & $0.45\pm0.04$ & \tabularnewline
 & $0.60$ & $0.43\pm0.04$ & \tabularnewline
 & $0.78$ & $0.38\pm0.04$ &
\\\hline
2MTF & $0.02$ & $0.34\pm 0.04$ & Springob et al. 2016
\\\hline
VIPERS & $0.7$ & $0.38^{+0.06}_{-0.07}$ &  Granett et al. 2015
\\\hline
VVDS & $0.77$ & $0.49\pm0.18$ & Guzzo et al. 2008; Song \& Percival 2009
\\\hline
 \end{tabular}
 }
\caption{Published values of $f(z)\sigma_8(z)$, referred to in the text as the LSS formation data.}
\end{center}
\end{table}
%
\section{Structure formation with dynamical vacuum}\label{sect:perturbationsDVM}

Despite the theory of cosmological perturbations has been discussed at length in several specialized textbooks, see e.g. (Peebles 1993; Liddle \& Lyth 2000, 2009; Dodelson 2003), the dynamical character of the vacuum produces some changes on the standard equations which are worth mentioning.  At deep subhorizon scales one can show that the matter density contrast $\delta_m=\delta\rho_m/\rho_m$ obeys the following differential equation (cf. (Basilakos \& Sol\`a 2014; G\'omez-Valent, Sol\`a \& Basilakos 2015; G\'omez-Valent \& Sol\`a 2018) for details):
\begin{equation}\label{diffeqD}
\ddot{\delta}_m+\left(2H+\Psi\right)\,\dot{\delta}_m-\left(4\pi
G\rmr-2H\Psi-\dot{\Psi}\right)\,\delta_m=0\,,
\end{equation}
where $\Psi\equiv -\dot{\rho}_{\Lambda}/{\rmr}= Q/{\rmr}$, and the (vacuum-matter) interaction source $Q$ for each DVM is given in Sect.\,\ref{sect:DVMs}. For $\rL=$const. and for the XCDM and CPL there is no such an interaction and Eq.\,(\ref{diffeqD}) reduces to
$\ddot{\delta}_m+2H\,\dot{\delta}_m-4\pi G\rmr\,\delta_m=0$, i.e. the $\CC$CDM form (Peebles 1993).
We note that at the scales under consideration we are neglecting the perturbations of the vacuum energy density in front of the perturbations of the matter field. The justification for this has recently been analyzed in detail, cf. (G\'omez-Valent \& Sol\`a 2018).

Let us briefly justify by two alternative methods the modified form (\ref{diffeqD}), in which the variation of $\rho_\CC$ enters  through the Hubble function and the background quantity $\Psi$, but not through any perturbed quantity. We shall conveniently argue in the context of two well-known gauges, the synchronous gauge and the Newtonian conformal gauge, thus providing a twofold justification. In the synchronous gauge, vacuum perturbations $\delta\rL$ modify the momentum conservation equation for the matter particles in a way that we can easily get e.g. from the general formulae of (G\'omez-Valent \& Sol\`a 2018; Grande, Pelinson \& Sol\`a 2008), with the result
\begin{equation}\label{eq:ModifiedMomentum}
\dot{v}_m+ \, H v_m=\frac{1}{a}\frac{\delta\rho_\Lambda}{\rho_m}-\Psi v_m\,,
\end{equation}
where  $\vec{v}=\vec{\nabla}v_m$ is the associated peculiar velocity, with potential $v_m$ (notice that this quantity has dimension of inverse energy in natural units). By setting $\delta\rho_\Lambda= a\,Q\,{v_m}=a\,\rho_m\,\Psi\,v_m$ the two terms on the r.h.s. of (\ref{eq:ModifiedMomentum}) cancel each other and we recover the corresponding equation of the $\CC$CDM.
On the other hand, in the Newtonian or conformal gauge (Mukhanov, Feldman \& Brandenberger 1992; Ma \& Bertschinger 1995) we find a similar situation. The analog of the previous equation is the modified Euler's equation in the presence of dynamical vacuum energy, 
\begin{equation}\label{eq:Euler}
\frac{d}{d\eta}\left(\rho_mv_m\right)+4\mathcal{H}\rho_mv_m+\rho_m\phi-\delta\rho_\Lambda=0\,,
\end{equation}
where $\phi$ is the gravitational potential that appears explicitly in the Newtonian conformal gauge, and $\eta$ is the conformal time. Let an overhead circle denote a derivative with respect to the conformal time, $\mathring{f}=df/d\eta$ for any $f$. We define the quantities $\mathcal{H}=\mathring{a}/a=aH$ and $\bar{\Psi}=-\mathring{\rho}_{\Lambda}/\rho_m=a\Psi$, which are the analogues of $H$ and $\Psi$ in conformal time. Using the background local conservation equation (\ref{BianchiGeneral}) for the current epoch (neglecting therefore radiation) and rephrasing it in conformal time, i.e. $\mathring{\rho}_\Lambda+\mathring{\rho}_m+3\mathcal{H}\rho_m=0$, we can bring (\ref{eq:Euler}) to 
\begin{equation}\label{eq:ModifiedEuler}
\mathring{v}_m+ \, \mathcal{H} v_m+\phi=\frac{\delta\rho_\Lambda}{\rho_m}-\bar{\Psi} v_m\,.
\end{equation}
Once more the usual fluid equation (in this case Euler's equation) is retrieved if we arrange that $\delta\rho_\Lambda=\rho_m\,\bar{\Psi}\,v_m=a\,\rho_m\,\Psi\,v_m$, as then the two terms on the r.h.s. of (\ref{eq:ModifiedEuler}) cancel each other. Alternatively, one can use the covariant form $\nabla^{\mu} T_{\mu\nu}=Q_\nu$ for the local conservation law, with  the source 4-vector $Q_\nu= Q U_\nu$, where $U_\nu=(a,\vec{0})$ is the background matter 4-velocity in conformal time. By perturbing the covariant conservation equation one finds

\be
\delta\left(\nabla^{\mu} T_{\mu\nu}\right)=\delta Q_\nu=\delta Q\,U_\nu+Q\delta U_\nu\,,
\ee
where $\delta Q$ and $\delta U_\nu=a(\phi,-\vec{v})$ are the perturbations of the source function and the 4-velocity, respectively. Thus, we obtain

\be
\delta\left(\nabla^{\mu} T_{\mu\nu}\right)=a(\delta Q+Q\phi,-Q\vec{v})\,.
\ee
From the $\nu=j$ component of the above equation, we derive anew the usual Euler equation $\mathring{v}_m+ \, \mathcal{H} v_m+\phi=0$, which means that the relation $\delta\rho_\Lambda= a Q\,{v_m}=a\,\rho_m\,\Psi\,v_m$ is automatically fulfilled. So the analyses in the two gauges converge to the same final result for $\delta\rL$.

After we have found the condition that $\delta\rho_\CC$ must satisfy in each gauge so as to prevent that the vacuum modifies basic conservation laws of the matter fluid, one can readily show that any of the above equations (\ref{eq:ModifiedMomentum}) or (\ref{eq:ModifiedEuler}) for each gauge (now with their r.h.s. set to zero), in combination with the corresponding perturbed continuity equation and the perturbed $00$-component of Einstein's equations (giving Poisson's equation in the Newtonian approximation), leads to the desired matter perturbations equation\,(\ref{diffeqD}), in accordance with the result previously derived by other means in Refs. (G\'omez-Valent, Sol\`a \& Basilakos 2015; Basilakos \& Sol\`a 2014). Altogether, the above considerations formulated in the context of different gauges allow us to consistently neglect the DE perturbations at scales down the horizon. This justifies the use of Eq.\,(\ref{diffeqD}) for the effective matter perturbations equation in our study of linear structure formation in the framework of the DVMs.  {See (G\'omez-Valent \& Sol\`a 2018) for an expanded exposition of these considerations.}

For later convenience let us also rewrite Eq.\,(\ref{diffeqD}) in terms of the scale factor variable rather than the cosmic time. Using $d/dt=aH\,d/da$ and denoting the differentiation  $d/da$ with a prime, we find:
\begin{equation}\label{diffeqDa}
\delta''_m + \frac{A(a)}{a}\delta'_m + \frac{B(a)}{a^2}\delta_m = 0\,,
\end{equation}
where the functions $A$ and $B$ of the scale factor are given by
\begin{align}
& A(a) = 3 + a\frac{H'(a)}{H(a)} + \frac{\Psi(a)}{H(a)}\,,\label{deffA}\\
& B(a) = - \frac{4\pi{G}\rho_m(a)}{H^2(a)} + \frac{2\Psi(a)}{H(a)} + a\frac{\Psi'(a)}{H(a)}\label{deffB}\,.
\end{align}

\begin{figure}
\centering
\includegraphics[width=0.9\linewidth]{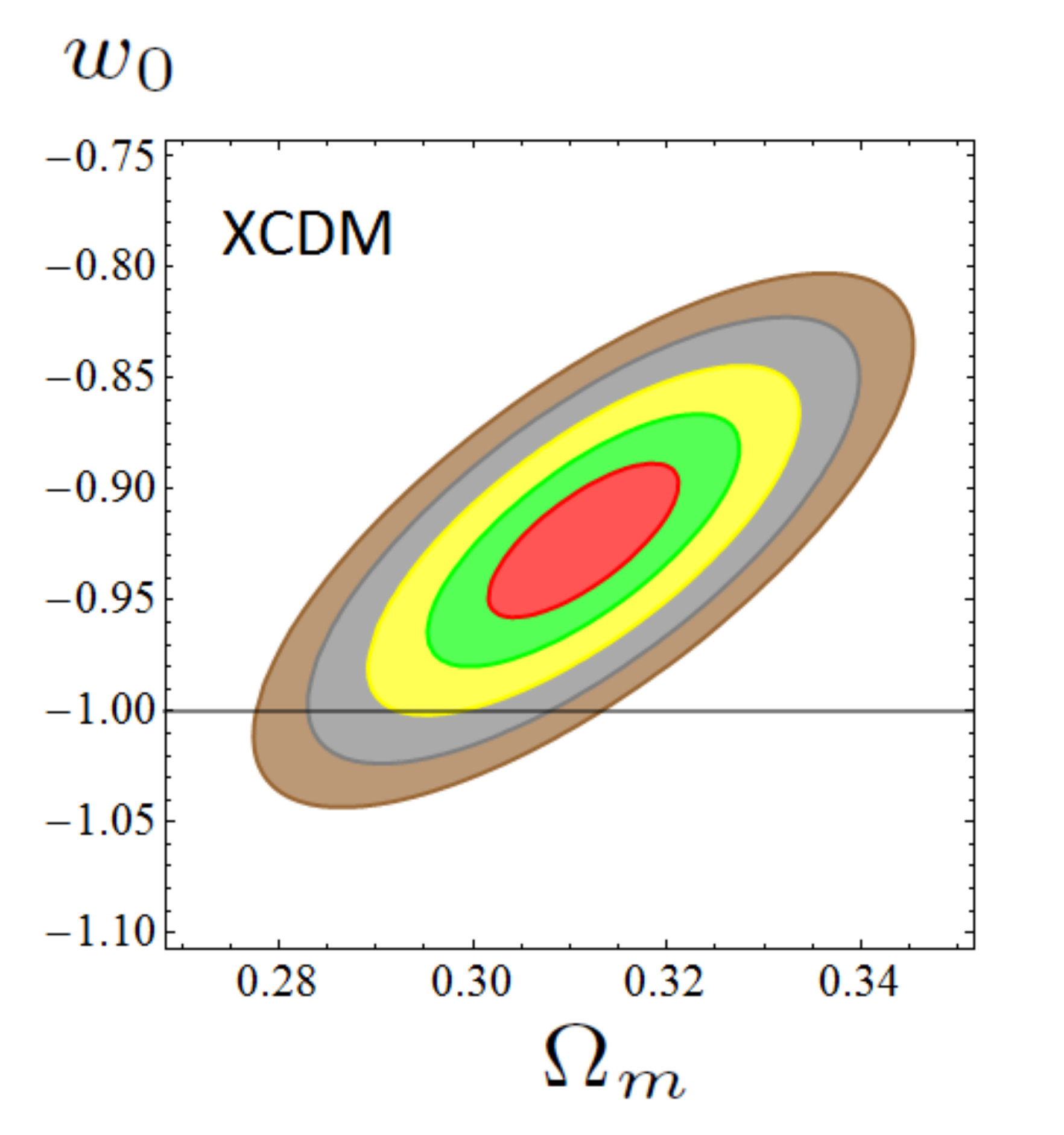}
\caption{\label{fig:XCDMEvolution}%
As in Fig.\,1, but for model XCDM. The $\CC$CDM is excluded in this case at $\sim 3\sigma$ c.l. Marginalization over $\Omega_m$ increases the c.l. up to $3.35\sigma$ (cf. Table 1).
}
\end{figure}
%

\subsection{Initial conditions}

{In order to solve \eqref{diffeqDa} we have to fix appropriate  initial conditions for $\delta_m(a)$ and $\delta'_m(a)$ for each model at high redshift, say at $z_i\sim100$ ($a_i\sim10^{-2}$), when nonrelativistic matter dominates both over the vacuum and the radiation contributions. In practice it  can be fixed at lower  redshifts, say of order $10$, where the subhorizon approximation is even more efficient (G\'omez-Valent \& Sol\`a 2018), although the differences are small. For small values of the scale factor the normalized Hubble rate (squared) for each model, and the energy densities for the various components,  see equations \eqref{eq:rhoRVM}-\eqref{HQL}, can be significantly simplified.
As a result we obtain the leading form of the functions \eqref{deffA}-\eqref{deffB} for the different DVMs:}
\begin{eqnarray}
{\rm RVM}:\phantom{X} A &=& \frac{3}{2}(1+3\nu) \label{AFunction1}\\
{\rm \ Q_{dm}}:\phantom{X} A &=& \frac{3}{2}\left(1 + \nu_{dm}\right) + 3\frac{\Omega_{dm}}{\Omega_m}\nu_{dm}+\mathcal{O}(\nu^2_{dm}) \nonumber\\ \label{AFunction2}\\
{\rm \ Q_{\CC}}:\phantom{X} A &=& \frac{3}{2}\,,
\label{AFunction3}
\end{eqnarray}
and
\begin{eqnarray}
{\rm RVM}:\phantom{X} B &=& -\frac{3}{2} +3\nu + \frac{9}{2}\nu^2  \label{BFunction1}\\
{\rm \ Q_{dm}}:\phantom{X} B &=& -\frac{3}{2}\left(1-\nu_{dm} - \frac{\Omega_{dm}}{\Omega_m}\nu_{dm}\right)+\mathcal{O}(\nu^2_{dm}) \nonumber\\ \label{BFunction2}\\
{\rm \ Q_{\CC}}: \phantom{X} B &=& -\frac{3}{2}\,.
\label{BFunction3}
\end{eqnarray}
For $\nu_i\rightarrow 0$, we recover the $\CC$CDM behavior $A \rightarrow \frac{3}{2}$ and $B \rightarrow -\frac{3}{2}$, as it should. This is already true for the $Q_{\CC}$ without imposing $\nu_{\CC}\rightarrow 0$, therefore its initial conditions are precisely the same as for the concordance model. Once the functions \eqref{deffA}-\eqref{deffB} take constant values (as it is the case here at the high redshifts where we fix the initial conditions), the differential equation \eqref{diffeqDa} admits power-like solutions of the form $\delta_m(a_i)=a_i^{s}$. Of the two solutions, we are interested only in the growing mode solution, as this is the only one relevant for structure formation. For example, using (\ref{AFunction1}) and (\ref{BFunction1}) for the case of the RVM, the perturbations equation (\ref{diffeqDa}) becomes
\begin{equation}\label{diffeqDaRVM}
\delta''_m + \frac{3}{2a}(1+3\nu)\delta'_m - \left(\frac{3}{2}-3\nu-\frac{9}{2}\,\nu^2\right)\frac{\delta_m}{a^2} = 0\,.
\end{equation}
The power-law solution for the growing mode gives the result $\delta_m=a^{1-3\nu}$, which is exact even keeping the ${\cal O}(\nu^2)$ term. Nevertheless, as warned previously, in practice we can neglect all ${\cal O}(\nu_i^2)$ contributions despite we indicate their presence.  Repeating the same procedure for the other models, the power-law behavior in each case for the growing mode solution $\delta_m\sim a^s$ is the following:
\begin{eqnarray}
{\rm RVM}:\phantom{X} s &=& 1-3\nu \label{svalues1}\\
{\rm \ Q_{dm}}:\phantom{X} s &=& 1-\nu_{dm}\left(\frac{6\Omega_m + 9\Omega_{dm}}{5\Omega_m}\right) + \mathcal{O}(\nu^2_{dm})\nonumber\\ \label{svalues2}\\
{\rm \ Q_{\CC}}:\phantom{X} s &=& 1\,.
\label{svalues3}
\end{eqnarray}
Imposing the above analytical results to fix the initial conditions, we are then able to solve numerically the full differential equation \eqref{diffeqDa} from a high redshift $z_i\sim100$ ($a_i\sim10^{-2}$) up to our days. The result does not significantly depend on the precise value of $z_i$, provided it is in the matter-dominated epoch and well below the decoupling time ($z\sim 10^3$), where the radiation component starts to be nonnegligible.

\subsection{Linear growth and growth index}

The linear growth rate of clustering is an important (dimensionless)
indicator of structure formation (Peebles 1993). It is defined as the logarithmic derivative of the linear growth factor $\delta_m(a)$ with respect to the log of the scale factor, $\ln a$. Therefore,
\begin{equation}\label{growingfactor}
f(a)\equiv \frac{a}{\delta_m}\frac{d\delta_m}{d a}=\frac{d{\rm ln}\delta_m}{d{\rm ln}a}\,,
\end{equation}
where $\delta_m(a)$ is obtained from solving
the differential equation (\ref{diffeqDa}) for each model. The physical significance of $f(a)$ is that it determines the peculiar velocity flows (Peebles 1993). In terms of the redshift variable, we have $f(z)=-(1+z)\,d{\ln\delta_m}/{dz}$, and thus the linear growth can also be used to determine the amplitude of the redshift distortions. This quantity has been analytically computed for the RVM in (Basilakos \& Sol\`a 2015). Here we shall take it into account for the study of the LSS data in our overall fit to the cosmological observations.

One usually expresses the linear growth rate of clustering
in terms of $\Omega_m(z)=\rho_m(z)/\rho_c(z)$, where $\rho_c(z)=3H^2(z)/(8\pi G)$ is the evolving critical density, as follows (Peebles 1993):
\begin{equation}\label{eq:gammaIndex}
f(z)\simeq \left[\Omega_{m}(z)\right]^{\gamma(z)}\,,
\end{equation}
where $\gamma$ is the so-called linear growth rate index. For the usual $\Lambda$CDM model, such an index is approximately given by $\gamma_{\CC} \simeq 6/11\simeq 0.545$. For models with a slowly varying  equation of state $w_D$ (i.e. approximately behaving as the XCDM, with $w_D\simeq w_0$) one finds the approximate formula $\gamma_D\simeq 3(w_D-1)/(6w_D-5)$ (Wang \& Steinhardt 1998) for the asymptotic value when $\Omega_m\to 1$. Setting $w_D=-1+\epsilon$, it can be rewritten
\begin{equation}\label{eq:GammaIndex}
\gamma_{D}\simeq \frac{6-3\epsilon}{11-6\epsilon}\simeq \frac{6}{11}\left( 1+\frac{1}{22}\,\epsilon\right)\,.
\end{equation}
Obviously, for $\epsilon\to 0$ (equivalently, $\omega_D\to -1$) one retrieves the $\CC$CDM case.
Since the current experimental error on the $\gamma$-index is of order $10\%$, it opens the possibility to discriminate cosmological models using such an index, see e.g. (Pouri, Basilakos \& Plionis 2014). In the case of the RVM and various models and frameworks, the function $\gamma(z)$ has been computed numerically in (G\'omez-Valent, Sol\`a \& Basilakos 2015). Under certain approximations, an analytical result can also be obtained for the asymptotic value (Basilakos \& Sol\`a 2015):
\begin{equation}\label{eq:GammaIndexRVM}
\gamma_{\rm RVM}\simeq \frac{6+3\nu}{11-12\nu}\simeq \frac{6}{11}\left(1+\frac{35}{22}\,\nu\right)\,.
\end{equation}
This expression for the RVM is similar to (\ref{eq:GammaIndex}) for an approximate XCDM parametrization, and it reduces to the $\CC$CDM value for $\nu=0$, as it should.

\subsection{Weighted linear growth and power spectrum}\label{sect:fsigma8}

A most convenient observable to assess the performance of our vacuum models in regard to structure formation is the combined quantity $f(z)\sigma_{8}(z)$, viz. the ordinary growth rate weighted with $\sigma_{8}(z)$, the rms total matter fluctuation
(baryons + CDM) on $R_8 = 8{h^{-1}}$ Mpc spheres at the given redshift $z$, computed in linear theory. It has long been recognized that this estimator is almost a model-independent way of expressing the observed growth history of the universe, most noticeably it is found to be independent of the galaxy density bias (Guzzo et al. 2008; Song \& Percival 2009).

With the help of the above generalized matter perturbations equation (\ref{diffeqDa}) and the appropriate initial conditions, the analysis of the linear LSS regime is implemented on using  the weighted linear growth $f(z)\sigma_8(z)$. The variance of the smoothed linear density field on $R_8 = 8{h^{-1}}$ Mpc spheres at redshift $z$ is computed from
\begin{equation}
\sigma_8^2(z)=\delta_m^2(z)\int\frac{d^3k}{(2\pi)^3}\, P(k,\vec{p})\,\,W^2(kR_8)\,.
\label{s88generalNN}
\end{equation}
Here ${P}(k,\vec{p})={P}_0\,k^{n_s}T^2(k)$ is the ordinary linear matter power spectrum (i.e. the coefficient of the two-point correlator of the linear perturbations), with $P_0$ a normalization factor, $n_s$ the spectral index and $T(k)$ the transfer function. Furthermore,
$W(kR_8)$ in the above formula is a top-hat smoothing function (see e.g. (G\'omez-Valent, Sol\`a \& Basilakos 2015) for
details), which can be expressed in terms of the spherical Bessel function of order $1$, as follows:
\begin{equation}\label{eq:WBessel}
W(kR_8)=3\,\frac{j_1(kR_8)}{kR_8}=\frac{3}{k^2R_8^2}\left(\frac{\sin{\left(kR_8\right)}}{kR_8}-\cos{\left(kR_8\right)}\right)\,.
\end{equation}
{Moreover, $\vec{p}$ is the  fitting vector with all the free parameters, including the specific vacuum parameters $\nu_i$ of the DVMs, or the EoS parameters $w_i$ for the XCDM/CPL parametrizations, as well as the standard parameters.}

{The power spectrum depends on all the components of the fitting vector. However, the dependence on the spectral index $n_s$ is  power-like, whereas the transfer function $T(k,\vec{q})$ depends in a  more complicated way on the rest of the fitting parameters (see below), and thus for convenience we collect them in the reduced fitting vector $\vec{q}$ not containing $n_s$.}
It is convenient to write the variance (\ref{s88generalNN}) in
terms of the dimensionless linear matter power spectrum, ${\cal P}(k,\vec{p})=\left(k^3/2\pi^2\right)\,P(k,\vec{p})$,
with
\begin{equation}\label{eq:PowerSpectrum}
{\cal P}(k,\vec{p})={\cal P}_0k^{n_s+3}T^2(k,\vec{q})\,.
\end{equation}
The normalization factor ${\cal P}_0=P_0/2\pi^2$  will be determined in the next section in connection to the definition of the  fiducial model.

For the transfer function, we have adopted the usual BBKS form (Bardeen, Bond, Kaiser \& Szalay 1986),
but we have checked that the use of the alternative
one by (Eisenstein \& Hu 1998) does not produce any significant change in our results. Recall that the wave number at equality, $k_{eq}$, enters the argument of the transfer function.
However,  $k_{eq}$ is a model-dependent quantity, which departs from the $\CC$CDM expression in those models in which matter and/or radiation are governed by an anomalous continuity equation, as e.g. in the DVMs. In point of fact $k_{eq}$ depends on all the parameters of the reduced fitting vector $\vec{q}$. For the concordance model, $k_{eq}$ has the simplest expression,
\begin{equation}\label{keqCCprev}
k^\CC_{eq} = H_0\,\Omega_m\sqrt{\frac{2}{\Omega_r}}=\frac{\Omega_mh^2}{2997.9}\sqrt{\frac{2}{\omega_r}}\,\textrm{Mpc}^{-1}\,,
\end{equation}
where $\omega_r=\Omega_r h^2$. In the second equality we have used the relation $H_0^{-1}=2997.9 h^{-1}$ Mpc.
For the DVMs it  is not possible to find a formula as compact as the one above. Either the corresponding expression for $a_{eq}$ is quite involved, as in the RVM case,
\begin{equation}\label{eq:aeqRVM}
\textrm{RVM}:\quad a_{eq}=\left[\frac{\Omega_r(1+7\nu)}{\Omega_m(1+3\nu)+4\nu\Omega_r}\right]^{\frac{1}{1+3\nu}}\,,
\end{equation}
or because $a_{eq}$ must be computed numerically, as for the models Q$_{dm}$ and Q$_\CC$. In all cases, for $\nu_i=0$ we retrieve the value of $a_{eq}$ in the $\CC$CDM.

\subsection{Fiducial model}\label{sect:FiducialModel}

Inserting  the dimensionless power spectrum \,(\ref{eq:PowerSpectrum}) into the variance (\ref{s88generalNN}) at $z=0$ allows us to write $\sigma_8(0)$ in terms of the power spectrum normalization factor ${\cal P}_0$ in (\ref{eq:PowerSpectrum}) and the primary parameters that enter our fit for each model. This is tantamount to saying that ${\cal P}_0$ can be fixed as follows:
\begin{equation}\label{P0}
\begin{small}
{\cal P}_0=\frac{\sigma_{8,\Lambda}^2}{\delta^2_{m,\Lambda}}\left[\int_{0}^{\infty} k^{n_{s,\Lambda+3}}T^2(k,\vec{q}_{\Lambda})W^2(kR_{8,\Lambda})(dk/k)\right]^{-1}\,,
\end{small}
\end{equation}
%
\begin{figure*}
\centering
\includegraphics[angle=0,width=0.7\linewidth]{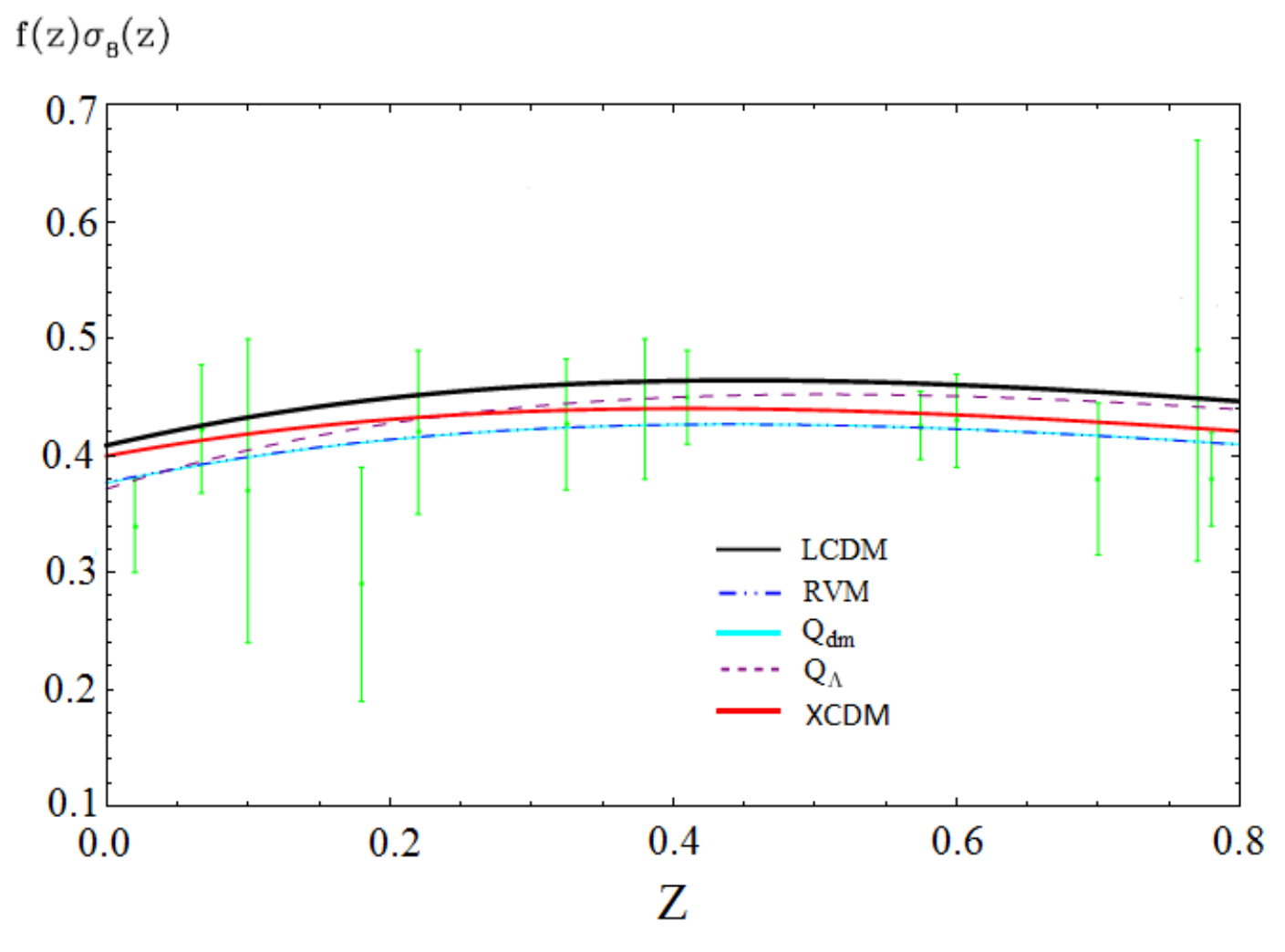}
\caption{\label{fig:fsigma8}%
The $f(z)\sigma_8(z)$ data (cf. Sect. \ref{sect:Fit}) and the predicted curves by the $\CC$CDM, XCDM, and the DVMs, for the best-fit values in Table 1.}
\end{figure*}
%

where the chosen values of the parameters in this expression define our fiducial model.  {The latter is characterized by the vectors of fiducial parameters $\vec{p}_\CC$
and  $\vec{q}_\CC$, defined in obvious analogy with the original fitting vectors but with all their parameters taken to be equal to those from the Planck 2015 TT,TE,EE+lowP+lensing analysis (Planck collab. XIII 2016), with $\nu_i=0$ for the DVMs and $w_0=-1, w_1=0$ for the XCDM/CPL parametrizations. The subindex $\CC$  carried by all the parameters denotes such a setting}.  In particular,  $\sigma_{8,\Lambda}\equiv\sigma_{8,\Lambda}(0)$  in (\ref{P0}) is also taken from the aforementioned Planck 2015 data.  However, $\delta_{m,\Lambda}\equiv\delta_{m,\Lambda}(0)$ in the same formula is computable: it is the value of $\delta_m(z=0)$ obtained from solving the perturbations equation of the $\CC$CDM, using the mentioned fiducial values of the other parameters. Finally, plugging the normalization factor \eqref{P0} in  (\ref{s88generalNN}) and using \eqref{eq:PowerSpectrum} one finds:
\begin{equation}
\begin{small}\label{eq:sigma8normalized}
\sigma_{8}(z)=\sigma_{8, \Lambda}
\frac{\delta_m(z)}{\delta_{m,\CC}}
\sqrt{\frac{\int_0^\infty k^{n_s+2} T^{2}(k,\vec{q})W^2(kR_{8})\, dk}{\int_0^\infty k^{n_{s,\CC}+2} T^{2}(k,\vec{q}_\Lambda) W^2(kR_{8,\Lambda})\, dk}}\,.
\end{small}\end{equation}
For the fiducial $\CC$CDM, this expression just gives the scaling of $\sigma_{8,\CC}(z)$ with the redshift in the linear theory, that is to say, $\sigma_{8,\CC}(z)/\sigma_{8,\CC}=\delta_{m,\CC}(z)/\delta_{m,\CC}$. But for an arbitrary model, Eq.\,(\ref{eq:sigma8normalized}) normalizes the corresponding $\sigma_{8}(z)$ with respect to the fiducial value, $\sigma_{8, \Lambda}$. This includes, of course, our fitted $\CC$CDM, which is not the same as the fiducial $\CC$CDM. {So all fitted models are compared to the same fiducial model defined by the Planck 2015 results.} Similarly, upon computing with this method the weighted linear growth rate  $f(z)\sigma_8(z)$ for each model under consideration, (including the $\CC$CDM) the functions $f(z)\sigma_8(z)$ for all models become normalized to the same fiducial model. It is important to emphasize that one cannot adjust the power spectrum and the $f(z)\sigma_8(z)$ values independently. Therefore, we first normalize with Planck 2015 results, as above described, and from here we fit the models to the data, in which the LSS  component takes an essential part.

The connection of the normalization factor (\ref{P0}) with $A_s$ (Planck collab. XIII 2016) can be easily found using standard formulae (Liddle \& Lyth 2000, 2009; Dodelson 2003). We find:
\begin{equation}\label{eq:PoAs}
{\cal P}_0=\frac{4A_s}{25}\frac{k_{*}^{1-n_s}}{H_0^4 \Omega_m^2}\,,
\end{equation}
where $k_{*}=0.05$ Mpc$^{-1}$ is the pivot scale used by Planck. This follows from the fact that ${\cal P}_0$ is related to $\delta_H^2$ (the primordial amplitude of the gravitational potential) through ${\cal P}_0={\delta_H^2}/(H_0^{3+n_s} \Omega_m^2)$ and on the other hand we have  $\delta_H^2=(4/25) A_s (H_0/k_{*})^{n_s-1}$.

In Fig.\,4 we display the theoretical results for $f(z)\sigma_8(z)$ from the various models, side by side with the LSS data measurements, using the fitted values of Table 1. {The values that we find for $\sigma_8(0)$ for each model, with the corresponding uncertainties, are reckoned in Table 1}.
Inspection of Fig.\,4 shows that the DVMs provide a better description of the LSS data points as compared to the $\CC$CDM. {The XCDM parametrization takes an intermediate position, granting a better fit than the $\CC$CDM but a poorer one than the RVM and $Q_{dm}$}. One can see that it is necessary an overall reduction of  $\sim 8\%$ in the value of  $f(z)\sigma_8(z)$  with respect to the $\CC$CDM curve (the solid line on top of the others in that figure).  Once $\Omega_m $ is accurately fixed from the CMB data, the $\CC$CDM model does not have any further freedom to further adjust the low-$z$ LSS data.  This can be seen from Eq.\,(\ref{eq:PoAs}) and from the fact that the normalization amplitude of the power spectrum $A_s$ as given by Planck tolerates an error of order $2\%$ at most(Planck collab. XIII 2016) and, therefore, such residual freedom cannot be invested to adjust the structure formation data, it is simply insufficient as we have checked. Thus, there seems to be no way at present to describe correctly  both the CMB and the LSS data within the  $\CC$CDM. This is of course at the root of the so-called  $\sigma_8$-tension, one of the important problems of the $\CC$CDM mentioned in the introduction -- see e.g. (Macaulay, Wehus \& Eriksen 2013; Battye, Charnock \& Moss 2015; Basilakos \& Nesseris 2016, 2017) for additional discussion and references.

In contrast, the DVMs can provide a possible clue.  For example, for  the RVM case an analytical explanation has recently been provided  in  Refs. (G\'omez-Valent \& Sol\`a 2017, 2018) showing why the dynamical vacuum can help in relaxing such tension. Recall that for $\nu=0$ the equality point between matter and radiation as given in Eq.\,(\ref{eq:aeqRVM}) boils down to the $\CC$CDM value. However, for $\nu\neq 0$ a nonnegligible contribution is obtained, despite the smallness of  $\nu$. Indeed, one can show that the $\nu$-effect causes a negative correction to the transfer function, which at linear order in $\nu$ is proportional to $6\nu\ln({\Omega_m}/{\Omega_r}) \simeq 50\,\nu$.  Since $\nu$ is fitted to be of order $\sim 10^{-3}$ in Table 1, it follows that the aforementioned negative correction can easily enhance the final effect up to near  $ 10\%$  level.  Upon a careful analysis of all the contributions, it eventually amounts to a  $\sim 8\%$ reduction of the weighted growth rate $f(z)\sigma_8(z)$ as compared to the $\CC$CDM value (G\'omez-Valent \& Sol`a 2017, 2018). This is precisely the reduction with respect to the $\CC$CDM prediction that is necessary in order to provide a much better description of the LSS data, see Fig.\,4. Interestingly enough, {as a bonus one also obtains an excellent description of the current weak-lensing data,  see  Sect.\,\ref{subsect:weaklensing}.}

\section{Main numerical results}\label{sect:numerical results}

For the statistical analysis, we define the joint likelihood function as the product of the likelihoods for all the data sets. Correspondingly, for Gaussian errors the total $\chi^2$ to be minimized reads:
\be\label{chi2s}
\chi^2_{tot}=\chi^2_{SNIa}+\chi^2_{BAO}+\chi^2_{H}+\chi^2_{LSS}+\chi^2_{CMB}\,.
\ee
Each one of these terms is defined in the standard way and they include the corresponding covariance matrices.


%
\begin{table*}
\begin{center}
\begin{scriptsize}
\begin{tabular}{| c | c |c | c | c | c | c |}
\multicolumn{1}{c}{Model} &  \multicolumn{1}{c}{$h$}  &  \multicolumn{1}{c}{$\Omega_m$}&  \multicolumn{1}{c}{{\small$\nu_i$}}  & \multicolumn{1}{c}{$w$} & \multicolumn{1}{c}{$\Delta{\rm AIC}$} & \multicolumn{1}{c}{$\Delta{\rm BIC}$}\vspace{0.5mm}
\\\hline
{\small $\CC$CDM} & $0.685\pm 0.004$ & $0.304\pm 0.005$ & - & -1 &  - & -\\
\hline
XCDM  &  $0.684\pm 0.009$&  $0.305\pm 0.007$& - & $-0.992\pm0.040$ &  -2.25 & -4.29 \\
\hline
RVM  & $0.684\pm 0.007$&  $0.304\pm 0.005$ & $0.00014\pm 0.00103$ & -1 &   -2.27 & -4.31 \\
\hline
$Q_{dm}$ &  $0.685\pm 0.007$&  $0.304\pm 0.005 $ & $0.00019\pm 0.00126 $ & -1 &   -2.27 & -4.31 \\
\hline
$Q_\CC$  &  $0.686\pm 0.004$&  $0.304\pm 0.005$ & $0.00090\pm 0.00330$ & -1 &   -2.21 & -4.25 \\
\hline
 \end{tabular}
\end{scriptsize}
\end{center}
\caption{Same as in Table 1, but removing the LSS data set from our fitting analysis.}
\label{tableFit4}
\end{table*}


Table 1 contains the main fitting results, whereas the other tables display complementary information.
We observe from Fig.\,1 that the vacuum parameters, $\nu$ and $\nu_{dm}$, are neatly projected non null and positive for the RVM and the Q$_{dm}$.  {In the particular case of the RVM,  Fig.\,2 displays in a nutshell our main results in all possible planes of the fitting parameter space.  Fig.\, 4, on the other hand, indicates that the XCDM is also sensitive to the DDE signal. In all cases the LSS data play an important role (cf. Fig.\, 4).  Focusing on the model that provides the best fit, namely the RVM,  Figs.\,5-6 reveal the clue to the main data sources responsible for the final results.} We will further comment on them in the next sections. Remarkably enough, the significance of this dynamical vacuum effect reaches up to about $\sim 3.8\sigma$ c.l. after marginalizing over the remaining parameters.

\subsection{Fitting the data with the XCDM and CPL parametrizations}\label{sect:XCDMandCPLnumerical}

Here we further elaborate on the results we have found by exploring now the possible time evolution of the DE in terms of the well-known XCDM and CPL parametrizations (introduced in Sect.\,\ref{sect:XCDMandCPL}). For the XCDM,  $w=w_0$ is the (constant) equation of state (EoS) parameter for X, whereas for the CPL there is also a dynamical component introduced by $w_1$, see Eq.\,(\ref{eq:CPL}).  The corresponding fitting results for the XCDM parametrization is included in all our tables, along with those for the DVMs and the $\CC$CDM. For the main Table 1, we also include the CPL fitting results. For example, reading off Table 1 we can see that the best-fit value for $w_0$ in the XCDM is
\begin{equation}\label{eq:woRVM}
w_0=-0.923\pm0.023.
\end{equation}
It is worth noticing that this EoS value is far from being compatible with a rigid $\CC$-term. It actually departs from $-1$ by precisely $3.35\sigma$ c.l. In Fig.\, 3 we depict the contour plot for the XCDM in the $(\Omega_m,w_0)$ plane. Practically all of the $3\sigma$-region lies above the horizontal line at $w_0=-1$. Subsequent marginalization over $\Omega_m$ renders the result (\ref{eq:woRVM}). Concerning the CPL, we can see from Table 1 that the errors on the fitting parameters are larger, specially on $w_1$, but it concurs with the XCDM that DE dynamics is also preferred (see also Sect.\,\ref{subsect:AICandBIC}).

\begin{small}
\begin{table*}
\begin{center}
\begin{scriptsize}
\begin{tabular}{| c | c |c | c | c | c | c | c | c | c | c|}
\multicolumn{1}{c}{Model} &  \multicolumn{1}{c}{$h$} &    \multicolumn{1}{c}{$\Omega_m$}&  \multicolumn{1}{c}{{\small$\nu_i$}}  & \multicolumn{1}{c}{$w$}  & \multicolumn{1}{c}{$\Delta{\rm AIC}$} & \multicolumn{1}{c}{$\Delta{\rm BIC}$}\vspace{0.5mm}
\\\hline
{\small $\CC$CDM} & $0.679\pm 0.005$ &  $0.291\pm 0.005$ & - & -1  &  - & -\\
\hline
XCDM  &  $0.674\pm 0.007$&  $0.298\pm 0.009$& - & $-0.960\pm0.038$   & -1.18 & -3.40 \\
\hline
RVM  & $0.677\pm 0.008$&  $0.296\pm 0.015$ & $0.00061\pm 0.00158$ & -1  & -2.10 & -4.32 \\
\hline
$Q_{dm}$ &  $0.677\pm 0.008$&  $0.296\pm 0.015 $ & $0.00086\pm 0.00228 $ & -1    & -2.10 & -4.32 \\
\hline
$Q_\CC$  &  $0.679\pm 0.005$&  $0.297\pm 0.013$ & $0.00463\pm 0.00922$ & -1   & -1.98 & -4.20 \\
\hline \end{tabular}

\end{scriptsize}
\end{center}
\caption{Same as in Table 1 but removing the CMB data set from our  fitting analysis.}
\label{tableFitextra}
\end{table*}
\end{small}


\begin{table*}
\begin{center}
\begin{scriptsize}
\begin{tabular}{| c | c |c | c | c | c | c |}
\multicolumn{1}{c}{Model} &  \multicolumn{1}{c}{$h$} &    \multicolumn{1}{c}{$\Omega_m$}&  \multicolumn{1}{c}{{\small$\nu_i$}}  & \multicolumn{1}{c}{$w$} &\multicolumn{1}{c}{$\Delta{\rm AIC}$} & \multicolumn{1}{c}{$\Delta{\rm BIC}$}\vspace{0.5mm}
\\\hline
{\small $\CC$CDM} & $0.694\pm 0.005$ &  $0.293\pm 0.007$ & - & -1 &   - & -\\
\hline
XCDM  &  $0.684\pm 0.010$&  $0.299\pm 0.009$& - & $-0.961\pm0.033$ &   -1.20 & -2.39 \\
\hline
RVM  & $0.685\pm 0.009$&  $0.297\pm 0.008$ & $0.00080\pm 0.00062$ & -1 &   -0.88 & -2.07 \\
\hline
$Q_{dm}$ &  $0.686\pm 0.008$&  $0.297\pm 0.008 $ & $0.00108\pm 0.00088 $ & -1 &   -1.02 & -2.21\\
\hline
$Q_\CC$  &  $0.694\pm 0.006$&  $0.293\pm 0.007$ & $0.00167\pm 0.00471$ & -1 &  -2.45 & -3.64 \\
\hline
 \end{tabular}
\end{scriptsize}
\end{center}
\caption{As in Table 1, but using the same data set as the Planck Collaboration (Planck collab. XIV 2016). }
\label{tableFit6}
\end{table*}

{Remarkably, from the rich string of SNIa+BAO+$H(z)$+LSS+CMB data we find that even the simple XCDM parametrization is able to capture nontrivial signs of dynamical DE in the form of effective quintessence behavior ($w_0\gtrsim -1$), at more than $3\sigma$ c.l.} Given the significance of this fact, it is convenient to compare it with well-known previous fitting analyses of the XCDM parametrization, such as the ones performed by the Planck and BOSS collaborations 2-3 years ago. The Planck 2015 value for the EoS parameter of the XCDM reads $w_0 = -1.019^{+0.075}_{-0.080}$ (Planck collab. XIII 2016) and the BOSS one is $w_0 = -0.97\pm 0.05$ (Aubourg et al. 2015). These results are perfectly compatible with our own fitting value for $w_0$ given in (\ref{eq:woRVM}), but in stark contrast to it their errors are big enough as to be also fully compatible with the $\CC$CDM value $w_0=-1$. This is not too surprising if we bear in mind that none of these analyses included large scale structure formation data in their fits, as explicitly recognized in the text of their papers.

In the absence of the modern LSS data we would indeed find a very different situation to that in Table 1. As our Table 3 clearly shows, the removal of the LSS data set in our fit induces a significant increase in the magnitude of the central value of the EoS parameter for the XCDM, as well as of the corresponding error. This happens because the higher is $|w|$ the higher is the structure formation power predicted by the XCDM, and therefore the closer is such a prediction with that of the $\CC$CDM (which is seen to predict too much power as compared to the data, see Fig.\,4). Under these conditions our analysis renders $w = -0.992\pm 0.040$ (cf. Table 3), which is manifestly closer to (in fact consistent with) the aforementioned central values (and errors) obtained by Planck and BOSS teams. In addition, this result is now fully compatible with the $\CC$CDM, as in the Planck 2015 and BOSS cases, and all of them are unfavored by the LSS data.

From the foregoing observations it becomes clear that in order to improve the fit to the observed values of $f(z)\sigma_8(z)$, which generally appear lower-powered with respect to those predicted by the $\CC$CDM (cf. Fig.\,4), $|w|$ should decrease. This is just what happens in our fit for the XCDM, see Eq.\,(\ref{eq:woRVM}).  At the level of the DVMs this translates into positive values of $\nu_i$, as these values cause the vacuum energy to be larger in our past; and, consequently, it introduces a time modulation of the growth suppression of matter. It is apparent from Fig.\,4 that the $f(z)\sigma_8(z)$ curves for the vacuum models are shifted downwards (they have less power than the $\CC$CDM) and hence adapt significantly better to the LSS data points.

%
\begin{figure*}
\centering
\includegraphics[angle=0,width=1.03\linewidth]{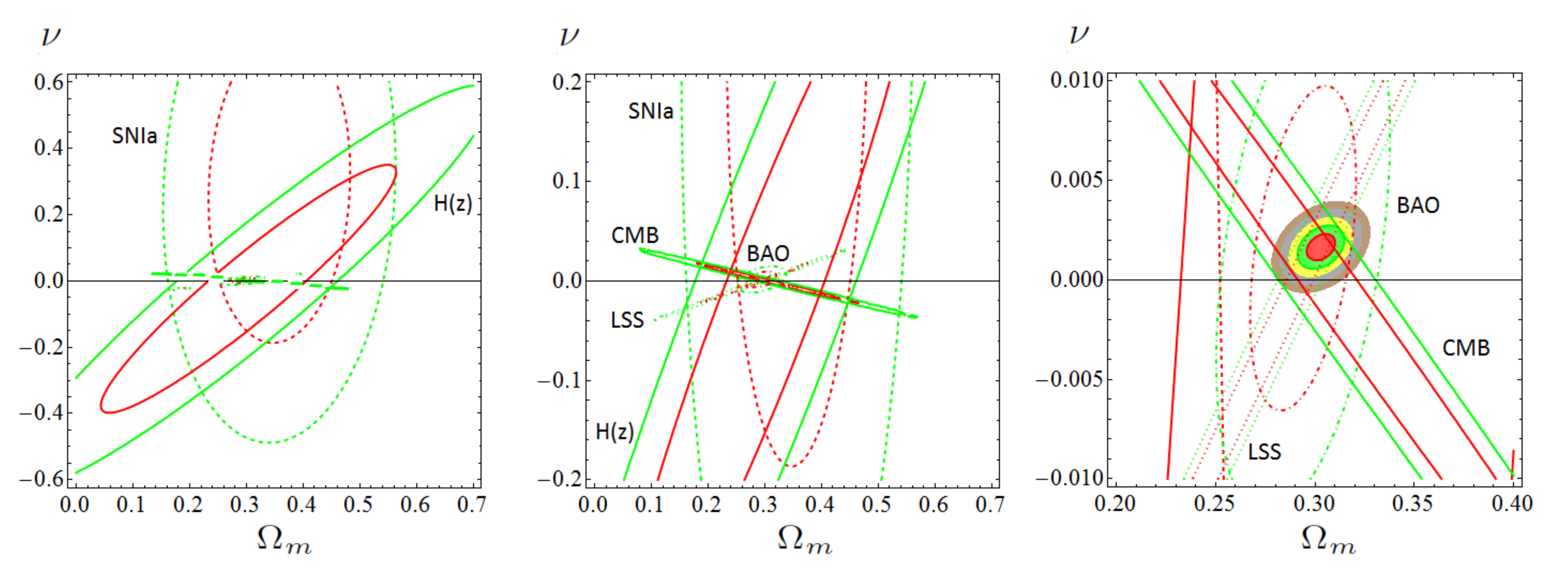}
\caption{Reconstruction of the contour lines for the RVM, from the partial contour plots of the different SNIa+BAO+$H(z)$+LSS+CMB data sources. The $1\sigma$ c.l. and $2\sigma$ c.l. contours are shown in all cases. For the reconstructed final contour lines we also plot the $3\sigma$, $4\sigma$ and $5\sigma$ confidence level regions.
}
\end{figure*}


\subsection{Comparing the competing vacuum models through Akaike and Bayesian information criteria}\label{subsect:AICandBIC}

We may judge the fit quality obtained for the different vacuum models in this work from a different perspective. Although the $\chi^2_{\rm min}$ value of the overall fits for the main DVMs (RVM and $Q_{dm}$) and XCDM appear to be definitely smaller than the $\CC$CDM one, it proves extremely useful to reassess the degree of success of each competing model by invoking the time-honored Akaike and Bayesian information criteria, denoted as AIC and BIC (Akaike 1974; Schwarz 1978; Kass \& Raftery 1995).
The Akaike information criterion is defined as follows:
\begin{equation}\label{eq:AIC}
{\rm AIC}=\chi^2_{\rm min}+\frac{2nN}{N-n-1}\,,
\end{equation}
whereas the Bayesian information criterion reads
\begin{equation}\label{eq:BIC}
{\rm BIC}=\chi^2_{\rm min}+n\,\ln N\,.
\end{equation}
In these formulas, $n$ is the number of independent fitting parameters and $N$ the number of data points. The added terms on $\chi^2_{\rm min}$ represent the penalty assigned by these information criteria to the models owing to the presence of additional parameters.
To test the degree of success of a dynamical DE model  (versus the $\CC$CDM) with the information criteria, we have to evaluate the pairwise differences $\Delta$AIC ($\Delta$BIC) between the AIC and BIC values of the $\CC$CDM with respect to the corresponding values of the models having a smaller value of these criteria -- in our case the DVMs, XCDM and CPL. The larger these (positive) differences are the higher is the evidence against the model with larger value of  AIC (BIC) -- i.e. the $\CC$CDM in the present case.

According to the standard usage, for $\Delta$AIC and/or $\Delta$BIC below 2 one judges that there is ``consistency'' between the two models under comparison; in the range $2-6$ there exists a ``positive evidence'' in favor of the model with smaller value of AIC and/or BIC; for values within $6-10$ one may claim ``strong evidence'' in favor of such a model; finally, above 10, one speaks of ``very strong evidence''. The evidence ratio associated to acceptance of the favored model and rejection of the unfavored model is given by the ratio of Akaike weights, $A\equiv e^{\Delta{\rm AIC}/2}$. Similarly, $B\equiv e^{\Delta{\rm BIC}/2}$ estimates the so-called Bayes factor, which gives the ratio of marginal likelihoods between the two models (Amendola 2015; Amendola \& Tsujikawa 2015).
{Table 1 reveals conspicuously that the $\CC$CDM appears disfavored when confronted to  the DDE models. The most favored one is the RVM, followed by the $Q_{dm}$ and next by the XCDM.  In the case of the CPL and  Q$_{\CC}$  the improvement is only mild.}

The AIC and BIC criteria can be thought of as a modern quantitative formulation of Occam's razor, in which the presence of extra parameters in a given model is conveniently penalized so as to achieve a fairer comparison with the model having less parameters.

\section{Discussion}\label{sect:discussion}

In this section we consider in more detail some important aspects and applications of our analysis. In particular we identify which are the most important data sources which are responsible for the possible DDE signal and show that in the absence of any of these important ingredients the signal becomes weakened or completely inaccessible.

\subsection{Testing the impact of the different data sets in our analysis and comparing with Planck 2015}\label{sect:OtherDataSets}

The current work follows the track of (Sol\`a, G\'omez-Valent \& de Cruz P\'erez 2015) and is also firmly aligned with (Sol\`a, G\'omez-Valent \& de Cruz P\'erez 2017a and  Sol\`a, de Cruz P\'erez  \& G\'omez-Valent  2018). Although the models analyzed in (Sol\`a, G\'omez-Valent \& de Cruz P\'erez 2015, 2017a) have some differences with respect to the ones treated here, the outcome of the analysis points to the very same direction, to wit: some DVMs  and the XCDM fit better the available data than the $\Lambda$CDM. But we want to emphasize some important aspects of the analysis carried out in this paper as compared to other analyses:

\begin{itemize}

\item We have used a large and fully updated set of cosmological  SNIa+BAO+$H(z)$+LSS+CMB observations. To our knowledge, this is one of the most complete and consistent data sets used in the literature, see (Sol\`a, G\'omez-Valent \& de Cruz P\'erez 2017a) up to some updating introduced here, specially concerning the LSS data.

\item We have removed all data that would entail double counting and used the known covariance matrices in the literature. As an example, we have avoided to use Hubble parameter data extracted from BAO measurements, {and restricted only to those based on the differential age (i.e. the cosmic chronometers).}

    \item We have duly taken into account all the known covariance matrices in the total $\chi^2$-function \eqref{chi2s}, which means that we have accounted for all the known correlations among the data.  Not all data sets existing in the literature are fully consistent, sometimes they are affected from important correlations that have not been evaluated. We have discussed the consistency of the present data in (Sol\`a, G\'omez-Valent \& de Cruz P\'erez 2017a).

\end{itemize}

{We have conducted several practical tests in order to study the influence of different data sets in our fitting analysis. As previously mentioned, we have checked what is the impact on our results if we omit the use of the LSS data (cf. Table 3), but in our study we have also assessed what happens  if we disregard the  CMB data (cf. Table 4) while still keeping all the remaining observations. The purpose of this test is to illustrate once more the inherent $\sigma_8$-tension existing between the geometry data and the structure formation data.   In both cases, namely when we dispense with the LSS or the CMB data, we find that  for all the models under study the error bars for the fitted DDE parameters ($w_i, \nu_i$) become critically larger  (sometimes they increase a factor 2-4)  than those displayed in Table 1,  and as a consequence  these parameters become fully compatible with the $\CC$CDM values (in particular $\nu_i=0$ for the DVMs) within  $1\sigma$ c.l. or less, which is tantamount to saying that the DDE effect is washed out. At the same time, and in full accordance with the mentioned results, the $\Delta$AIC and $\Delta$BIC information criteria become negative, which means (according to our definition in Sect.\,\ref{subsect:AICandBIC})  that none of these DDE models fits better the data than the $\CC$CDM under these particular conditions.  These facts provide incontestable evidence of the strong constraining power of the LSS as well as of the CMB data, whether taken individually or in combination, and of their capability for narrowing down the allowed region in the parameter space. In the absence of either one of them, the $\Lambda$CDM model is preferred  over the DDE models, but only at the expense of ignoring the CMB input, or the LSS data, both of which are of course of utmost importance.  Thus, the concordance model is now able to fit the LSS data better only because it became free from the tight CMB constraint on $\Omega_m$, which enforced the latter to acquire a larger value. Without such constraint, a lower $\Omega_m$ value can be chosen by the fitting procedure,  what in turn enhances the agreement with the $f(z)\sigma_8(z)$ data points.  We have indeed checked that the reduction of $\Omega_m$ in the $\CC$CDM  directly translates into an $8.6\%$ lowering of $\sigma_8(0)$ with respect to the value shown in Table 1 for this model, namely we find that  $\sigma_8(0)$
 changes from  $0.801\pm 0.009$ (as indicated in Table 1) to  $ 0.731\pm 0.019 $  when the CMB data are not used. Such substantial decrease tends to optimize the fit of the LSS data, but only at the expense of ruining the fit to the CMB when these data are restored. This is, of course, the very meaning of the $\sigma_8$-tension, which cannot be overcome at the moment within the $\CC$CDM. }

{In stark contrast with the situation in the  $\CC$CDM,  when the vacuum is allowed to acquire a mild dynamical component the $\sigma_8$-tension can be dramatically  loosened, see  (G\'omez-Valent \& Sol\`a 2017, 2018)  for a detailed explanation. This can be seen immediately on comparing the current auxiliary tables 3 and 4 with the main Table 1. Recall that a positive vacuum energy suppresses the growth of structure formation and this is one of the reasons why the $\CC$CDM model is highly preferred to the CDM with $\CC=0$. Similarly, but at a finer and subtler  level of precision, a time modulation of the growth suppression through dynamical vacuum energy or in general DDE should further help in improving the adjustment of the LSS data.  In our case this is accomplished e.g. by the $\nu$-parameter of the RVM, which enables a dynamical modulation of the growth suppression through the $\sim \nu H^2$ component of the vacuum energy density -- cf. Eq.\,(\ref{eq:RVMvacuumdadensity}). The presence of this extra degree of freedom allows the DVMs to better adjust the LSS data without perturbing the requirements from the CMB data (which can therefore preserve the standard $\Omega_m$ value obtained by  Planck 2015). The fact that this  readjustment of the LSS data by a dynamical component in the vacuum energy  is possible is because the epoch of structure formation is very close to the epoch when the DE starts to dominate, which is far away from the epoch when the CMB was released,  and hence any new feature of the DE can play a significant role in the LSS formation epoch without disrupting the main features of the  CMB.  Let us recall at this point that the presence of the extra parameter from the DDE models under discussion is conveniently penalized by the Akaike and Bayesian information criteria in our analysis, and thus the DDE models appear to produce a better fit than the $\CC$CDM under perfectly fair conditions of statistical comparison between competing models describing the same data.}

{The conclusion of our analysis is clear: no signal of DDE can be found without the inclusion of the CMB data and/or the LSS data, even keeping the rest of observables within the fit. Both the LSS and CMB data are crucial ingredients to enable capturing the DDE effect, and the presence of BAO data just enhances it further.  This conclusion is  additionally confirmed by our study of the deconstruction and reconstruction of the RVM contour plots  in Figs. 5-6 and is discussed at length  in the next section. }

\begin{figure*}
\centering
\includegraphics[angle=0,width=0.6\linewidth]{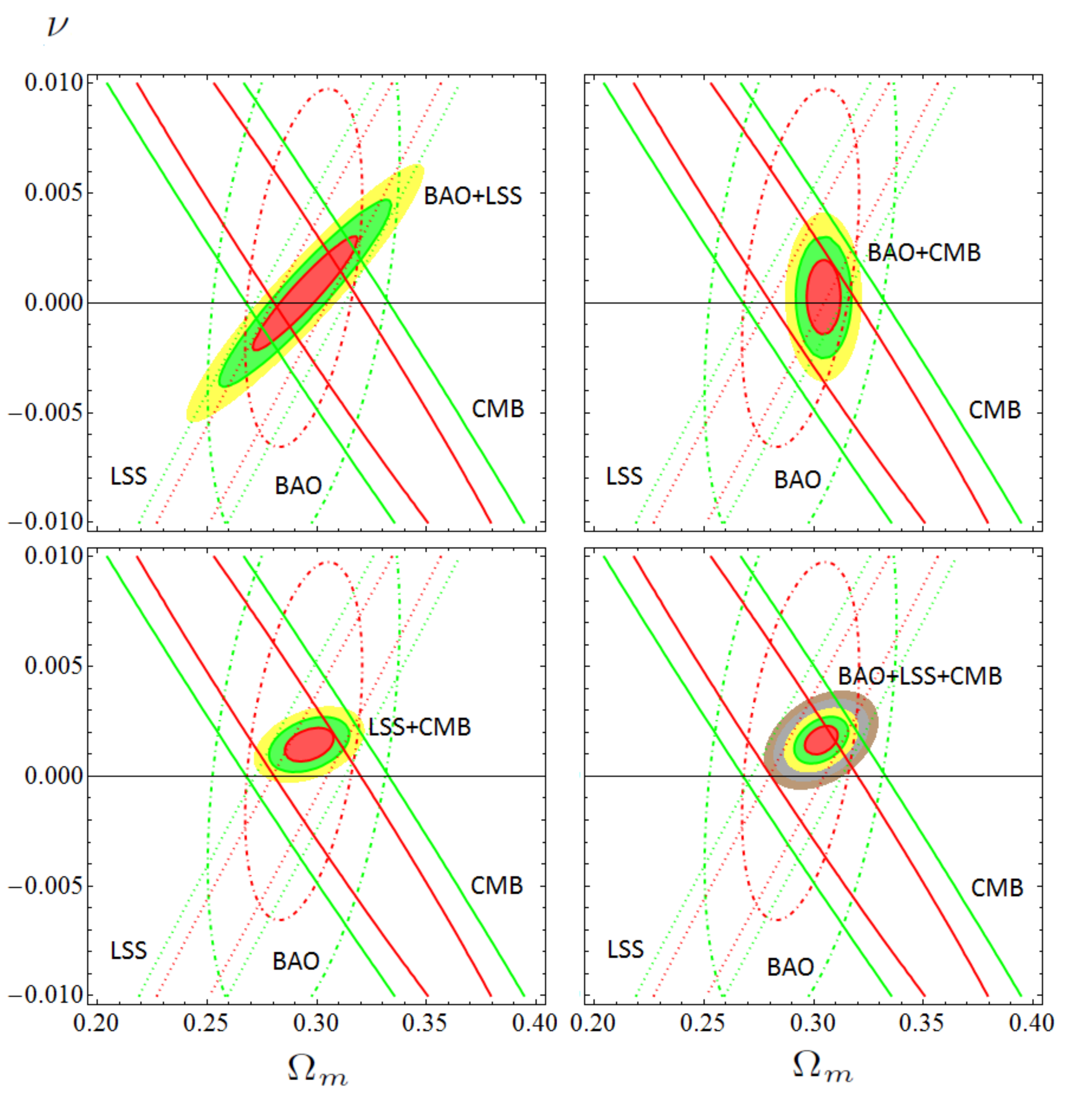}
\caption{\label{RecRVMtriad}%
As in Fig.\,5, but considering the effect of only the BAO, LSS and CMB in all the possible combinations: BAO+LSS, BAO+CMB, LSS+CMB and BAO+LSS+CMB. As discussed in the text, it is only when such a triad of observables is combined that we can see a clear $\lesssim 4\sigma$ c.l. effect, which is comparable to intersecting the whole set of SNIa+BAO+$H(z)$+LSS+CMB data.}
\end{figure*}


We close this section by answering a most natural question. Why the dynamical DE signal that we are glimpsing here escaped undetected from the fitting analyses of Planck 2015?  The answer can be obtained by repeating our fitting procedure and restricting ourselves to the much more limited data sets used by the Planck 2015 collaboration, more precisely in the papers (Planck collab. XIII 2016; Planck collab. XIV 2016). In contrast to (Planck collab. XIII 2016), where no LSS (RSD) data were used, in the case of (Planck collab. XIV 2016) they used only some BAO and LSS data, but their fit is rather limited in scope. Specifically, they used only 4 BAO data points, 1 AP (Alcock-Paczynski parameter) data point, and one single LSS point, namely the value of $f(z)\sigma_8(z)$ at $z=0.57$-- see details in that paper. Using this same data we obtain the fitting results presented in our Table 5. They are perfectly compatible with the fitting results mentioned in Sect.\,\ref{sect:XCDMandCPLnumerical} obtained by Planck 2015 and BOSS (Aubourg et al. 2015), i.e. none of them carries evidence of dynamical DE, with only the data used by these collaborations two-three years ago.

In contradistinction to them, in our full analysis presented in Table 1  we used 11 BAO and 13 LSS data points, some of them available only from the recent literature and of high precision (Gil-Mar\'in et al. 2017). From Table 5 it is apparent that with only the data used in (Planck collab. XIV 2016) the fitting results for the RVM are poor enough and cannot still detect clear traces of the vacuum dynamics. In fact, the vacuum parameters are compatible with zero at $1\sigma$ c.l. and the values of $\Delta$AIC and $\Delta$BIC in that table are moderately negative, showing that the DVMs do not fit better the data than the $\Lambda$CDM model with only such a limited input.  In fact, not even the XCDM parametrization is capable of detecting any trace of dynamical DE with that limited data set, as the effective EoS parameter is compatible with $w_0=-1$ at roughly $1\sigma$ c.l. ($w_0=-0.961\pm 0.033$).

The features that we are reporting here have remained hitherto unnoticed in the literature, except in (Sol\`a, G\'omez-Valent \& de Cruz P\'erez 2015, 2017a,b,c,d), and in (Zhao G-B. et al. 2017). In the last reference the authors have been able to find a significant $3.5\sigma$ c.l. effect on dynamical DE, presumably in a model-independent way and following a nonparametric procedure, see also (Wang, Zhao G-B., Wands, Pogosian \& Crittenden 2015). The result of (Zhao G-B et al. 2017) is well along the lines of the present work.

\begin{figure*}
\centering
\includegraphics[angle=0,width=0.75\linewidth]{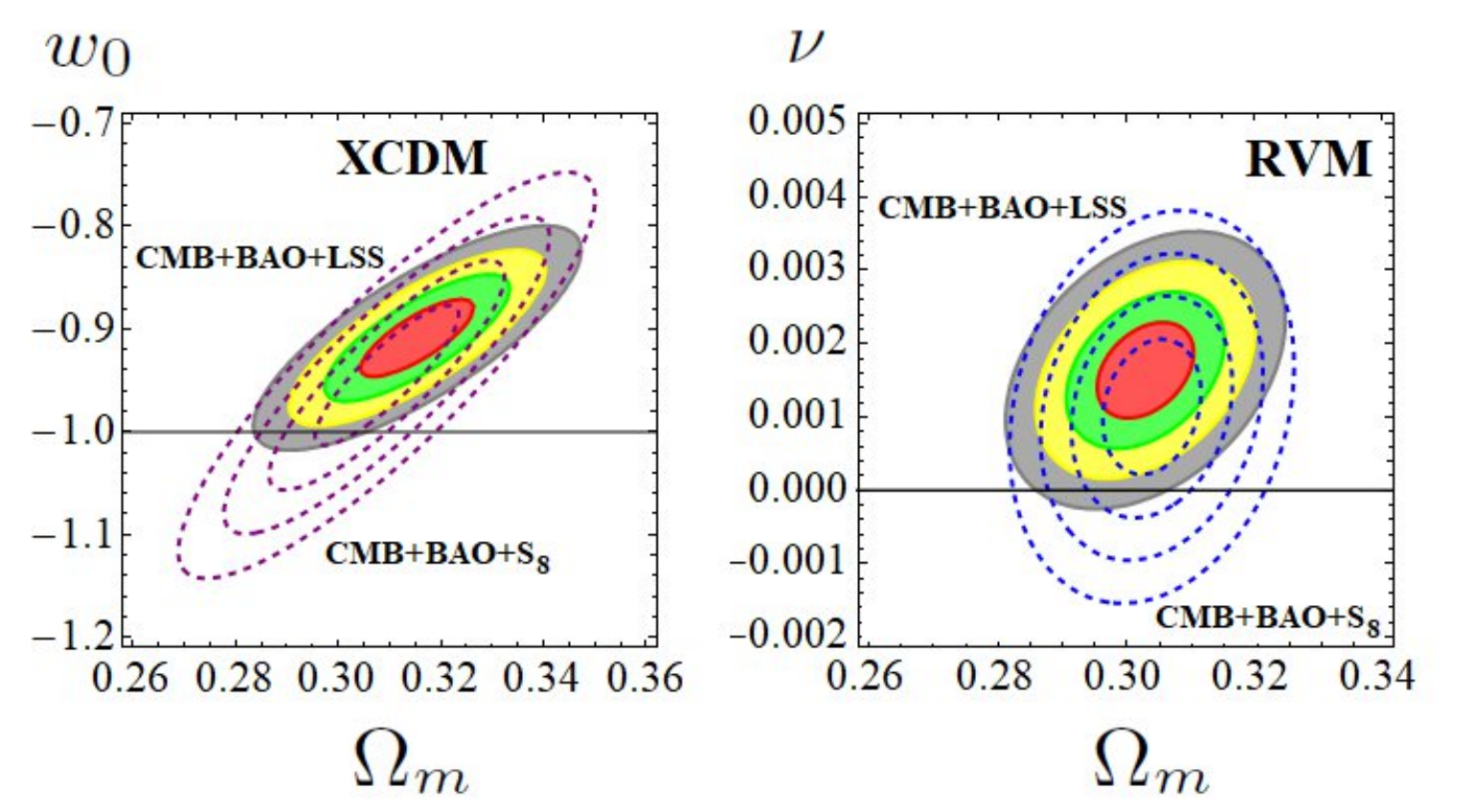}
\caption{\label{RecRVMtriad}%
Contour lines for the XCDM (left) and RVM (right) using the same CMB+BAO+LSS data as
in Table 1 (solid contours); and also when replacing the LSS data (i.e. the $f(z)\sigma_8(z)$ points) with the $S_8$ value obtained from the weak gravitational lensing data (Joudaki et al. 2018) (dashed lines).}
\end{figure*}


\subsection{Deconstruction and reconstruction of the RVM contour plots}\label{subsect:deconstruction}

We further complement our analysis by displaying in a graphical way the deconstructed contributions from the different data sets to our final contour plots in Fig.\,1, for the specific case of the RVM. One can do similarly for any of the models under consideration. The result is depicted in Fig.\,5, where we can assess the detailed deconstruction of the final contours in  terms of the partial contours from the different SNIa+BAO+$H(z)$+LSS+CMB data sources.

The deconstruction plot for the RVM case is dealt with in Fig.\,5, through a series of three plots made at different magnifications. In the third plot of the sequence we can immediately appraise that the BAO+LSS+CMB data subset plays a fundamental role in narrowing down the final physical region of the $(\Omega_m,\nu)$ parameter space, in which all the remaining parameters have been marginalized over. This deconstruction process also explains in very transparent visual terms why the conclusions that we are presenting here hinge to a large extent on some particularly sensitive components of the data. While there is no doubt that the CMB is a high precision component in the fit, our study demonstrates (both numerically and graphically) that the maximum power of the fit is achieved when it is combined with the wealth of BAO and LSS data points currently available.

To gauge the importance of the BAO+LSS+CMB combination more deeply, in Fig.\,6 we try to reconstruct the final RVM plot in Fig.\,1 (left) from only these three data sources. First we consider the overlapping regions obtained when we cross the pairs of data sources BAO+LSS, BAO+CMB, LSS+CMB and finally the trio BAO+LSS+CMB (in all cases excluding the SNIa and $H(z)$ data). One can see that neither the BAO+LSS nor the BAO+CMB crossings yield a definite sign for $\nu$. {This is consistent with the numerical results in Tables 3 and 4, where the removal of the LSS and the CMB data, respectively, renders rather poor fits with negative values of $\Delta$AIC and $\Delta$BIC.}

Remarkably, it is the LSS+CMB combination the one that carries a well-defined, positive, sign for $\nu$, as it is seen from the lower-left plot in Fig.\,6, where $\Delta$AIC and $\Delta$BIC are now both positive and above $6$ for the main DVMs (RVM and $Q_{dm}$), as we have checked. Finally, when we next intersect the pair LSS+CMB with the BAO data the signal peaks at  $3.8\sigma$ c.l., the final contours being now those shown in the lower-right plot of Fig.\,6.
The outcome of this exercise is clear. For the RVM case, we have checked that the final BAO+LSS+CMB plot in Fig.\,6 is essentially the same as the original RVM plot in Fig.\,1 (the leftmost one). In other words, the final RVM contour plot containing the information from all our five data sources can essentially be reconstructed with only the triad of leading observables BAO+LSS+CMB.

\subsection{Vacuum dynamics, structure formation and weak-lensing data}\label{subsect:weaklensing}

{Owing to the significant role played by the structure formation data in the extraction of the possible DDE signal we next  inquire into its impact when we use a different proxy to describe such data. Let us note that an account of the LSS observations does not only concern the $f(z)\sigma_8(z)$ data, but also the weak gravitational lensing constraints existing in the literature on the conventional quantity  $S_8\equiv \sigma_8(\Omega_m/0.3)^{0.5}$ (Joudaki et al. 2018; Hildebrandt et al. 2017; Heymans et al. 2013). In Fig.\,7 we compare the respective results that we find for the XCDM (left) and the RVM (right)  when we use either the CMB+BAO+$f\sigma_8$ or the CMB+BAO+$S_8$  data sources.  For definiteness we use the recent study by (Joudaki et al. 2018), in which they carry
a combined analysis of cosmic shear tomography, galaxy- galaxy lensing tomography, and redshift-space multipole
power spectra using imaging data by the Kilo Degree Survey (KiDS-450) overlapping with the 2-degree Field Lensing Survey (2dFLenS) and the Baryon Oscillation Spectroscopic Survey (BOSS). They find $S_8 = 0.742 \pm 0.035$. Incidentally, this value is $2.6\sigma$ below the one provided by Planck's TT+lowP analysis [4]. Very similar results can be obtained using the weak gravitational lensing tomography data by KiDS-450 collaboration, $S_8=0.745\pm 0.039$ (Hildebrandt et al. 2017), and also by CFHTLenS,  $(\Omega_m/0.27)^{0.46}=0.770\pm 0.040$ (Heymans et al. 2013).
In contrast, the result  $S_8=0.783^{+0.021}_{-0.025}$ provided by DES (DES collab. 2017) is more resonant with Planck, but due to its large uncertainty it is still fully compatible with (Joudaki et al. 2018; Hildebrandt et al. 2017; Heymans et al. 2013).  From Fig. 7 we confirm (using both the XCDM and the RVM) that the contour lines computed from the data string CMB+BAO+$f\sigma_8$  are mostly contained within the contour lines from the alternative string CMB+BAO+$S_8$ and are shifted upwards. The former data set is therefore more precise and capable of resolving the DDE signal at a level of more than $3\sigma$, whereas with $S_8$ it barely surpasses
the $1\sigma$  c.l. within the RVM and even less with the XCDM, thus rendering essentially no DDE signal. The outcome of this additional test is that the use of the weak-lensing data from $S_8$ as a replacement for the direct LSS measurements ($f\sigma_8$) is insufficient since it definitely weakens the evidence in favor of DDE.}

\section{Conclusions}\label{sect:conclusions}

To conclude, in this work we aimed at testing cosmological physics beyond the standard or concordance $\CC$CDM model, which is built upon a rigid cosmological constant. We have presented a comprehensive study on the possibility that the global cosmological observations can be better described in terms of vacuum models equipped with a dynamical component that evolves with the cosmic expansion. This should be considered a natural possibility in the context of quantum field theory (QFT) in a curved background. Our task focused on three dynamical vacuum models (DVMs): the running vacuum model (RVM) along with two more phenomenological models, denoted $Q_{dm}$ and $Q_\CC$-- see Sect.\,\ref{sect:DVMs}.

At the same time, we have compared the performance of these models with the general XCDM and CPL parametrizations. We have fitted all these models and parametrizations to the same set of cosmological data based on the observables SNIa+BAO+$H(z)$+LSS+CMB.
The remarkable outcome of this investigation is that in all the considered cases we find an improvement of the description of the cosmological data in comparison to the $\CC$CDM.

The ``deconstruction analysis'' of the contour plots in Sect.\,\ref{subsect:deconstruction} has revealed which are the most decisive data ingredients responsible for the dynamical vacuum signal. We have identified that the BAO+LSS+CMB components play a momentous role in the overall fit, as they are responsible for the main effects uncovered here. The impact of the SNIa and $H(z)$ observables appears to be more moderate.  While the SNIa data were of course essential for the detection of a nonvanishing value of $\CC$, these data do not seem to have sufficient sensitivity (at present) for the next-to-leading step, which is to unveil the possible dynamics of $\CC$. The sensitivity for that seems to be reserved for the LSS, BAO and CMB data.

We have also found that the possible signs of DDE tend to favor an effective quintessence behavior, in which the energy density decreases with the expansion. Whether or not the ultimate reason  for such a signal stems from the properties of the quantum vacuum or from some particular quintessence model, it is difficult to say at this point.
Quantitatively, the best fit is granted in terms of the RVM. The results are consistent with the traces of  DDE that can also be hinted at with the help of the XCDM and CPL  parametrizations.

In our work we have also clarified why previous fitting analyses based e.g. on the simple XCDM parametrization, such as the ones by the Planck 2015 (Planck collab. XIII 2015; Planck collab. XIV 2015) and BOSS collaborations (Aubourg et al. 2015), missed the DDE signature. Basically, the reason stems from not using a sufficiently rich sample of the most crucial data, namely BAO and LSS, some of which were unavailable a few years ago, and could not be subsequently combined with the CMB data.

More recently, signs  of DDE at $\sim 3.5\sigma$ c.l. have been reported from non-parametric studies of the observational data on the DE, which aim at a model-independent result (Zhao G-B et al. 2017). The findings of their analysis are compatible with the ones we have reported here.
Needless to say, statistical evidence conventionally starts at $5\sigma$ c.l. and we will have to wait for updated observations to see if such a level of significance can be achieved in the future.

\section{Acknowledgements}

We are partially supported by MINECO FPA2016-76005-C2-1-P,  2017-SGR-929 (Generalitat de Catalunya) and  MDM-2014-0369 (ICCUB).  JS acknowledges the hospitality and support received from the Institute for Advanced Study at the Nanyang Technological University in Singapore while part of this work was being accomplished.


\vspace{0.4cm}


\end{document}